\documentstyle[psfig]{mn}

\psfigurepath{.:/home/ciliegi/ELAIS/VLA/results/report/FIGURES}

\newif\ifAMStwofonts
\def\lsimeq{{_<\atop^{\sim}}}

\ifoldfss
  \ifCUPmtlplainloaded \else
    \NewTextAlphabet{textbfit} {cmbxti10} {}
    \NewTextAlphabet{textbfss} {cmssbx10} {}
    \NewMathAlphabet{mathbfit} {cmbxti10} {} % for math mode
    \NewMathAlphabet{mathbfss} {cmssbx10} {} %  "   "    "
  \fi
  \ifAMStwofonts
    \ifCUPmtlplainloaded \else
      \NewSymbolFont{upmath} {eurm10}
      \NewSymbolFont{AMSa} {msam10}
      \NewMathSymbol{\upi}     {0}{upmath}{19}
      \NewMathSymbol{\umu}     {0}{upmath}{16}
      \NewMathSymbol{\upartial}{0}{upmath}{40}
      \NewMathSymbol{\leqslant}{3}{AMSa}{36}
      \NewMathSymbol{\geqslant}{3}{AMSa}{3E}

      \let\geq=\geqslant 
    \fi
  \fi
\fi % End of OFSS

\ifnfssone
  \newmathalphabet{\mathit}
  \addtoversion{normal}{\mathit}{cmr}{m}{it}
  \addtoversion{bold}{\mathit}{cmr}{bx}{it}
  \newmathalphabet{\mathbfit} % math mode version of \textbfit{..}
  \addtoversion{normal}{\mathbfit}{cmr}{bx}{it}
  \addtoversion{bold}{\mathbfit}{cmr}{bx}{it}
  \newmathalphabet{\mathbfss} % math mode version of \textbfss{..}
  \addtoversion{normal}{\mathbfss}{cmss}{bx}{n}
  \addtoversion{bold}{\mathbfss}{cmss}{bx}{n}
  \ifAMStwofonts
    \ifCUPmtlplainloaded \else
      \UseAMStwoboldmath
      \makeatletter
      \new@mathgroup\upmath@group
      \define@mathgroup\mv@normal\upmath@group{eur}{m}{n}
      \define@mathgroup\mv@bold\upmath@group{eur}{b}{n}
      \edef\UPM{\hexnumber\upmath@group}
      \new@mathgroup\amsa@group
      \define@mathgroup\mv@normal\amsa@group{msa}{m}{n}
      \define@mathgroup\mv@bold\amsa@group{msa}{m}{n}
      \edef\AMSa{\hexnumber\amsa@group}
      \makeatother
      \mathchardef\upi="0\UPM19
      \mathchardef\umu="0\UPM16
      \mathchardef\upartial="0\UPM40
      \mathchardef\leqslant="3\AMSa36
      \mathchardef\geqslant="3\AMSa3E

      \let\geq=\geqslant 
    \fi
  \fi
\fi % End of NFSS release 1

\ifnfsstwo
  \DeclareMathAlphabet{\mathbfit}{OT1}{cmr}{bx}{it}
  \SetMathAlphabet\mathbfit{bold}{OT1}{cmr}{bx}{it}
  \DeclareMathAlphabet{\mathbfss}{OT1}{cmss}{bx}{n}
  \SetMathAlphabet\mathbfss{bold}{OT1}{cmss}{bx}{n}
  \ifAMStwofonts
    \ifCUPmtlplainloaded \else
      \DeclareSymbolFont{UPM}{U}{eur}{m}{n}
      \SetSymbolFont{UPM}{bold}{U}{eur}{b}{n}
      \DeclareSymbolFont{AMSa}{U}{msa}{m}{n}
      \DeclareMathSymbol{\upi}{0}{UPM}{"19}
      \DeclareMathSymbol{\umu}{0}{UPM}{"16}
      \DeclareMathSymbol{\upartial}{0}{UPM}{"40}
      \DeclareMathSymbol{\leqslant}{3}{AMSa}{"36}
      \DeclareMathSymbol{\geqslant}{3}{AMSa}{"3E}

      \let\geq=\geqslant 
    \fi
  \fi
\fi % End of NFSS release 2

\ifCUPmtlplainloaded \else
  \ifAMStwofonts \else % If no AMS fonts
    \def\upi{\pi}
    \def\umu{\mu}
    \def\upartial{\partial}
  \fi
\fi

\title{A Deep VLA survey at 20cm of the ISO ELAIS survey regions}
\author[P. Ciliegi et al.]
{\parbox[]{6.5in}
{P.~Ciliegi$^{1,2}$, 
R.G.~McMahon$^1$, 
G.~Miley$^3$,
C.~Gruppioni$^4$, 
M.~Rowan-Robinson$^4$,
C.~Cesarsky$^5$, 
L.~Danese$^6$,  
A.~Franceschini$^7$,
R.~Genzel$^8$,  
A.~Lawrence$^9$, 
D.~Lemke$^{10}$,  
S.~Oliver$^4$,  
J-L.~Puget$^{11}$,  
B.~Rocca-Volmerange$^{11,12}$ } \\ 
$^1$ Institute of Astronomy, University of Cambridge, Madingley Road, Cambridge CB3 0HA \\
$^2$ Osservatorio Astronomico di Bologna, via Zamboni 33, I-40126 Bologna, 
Italy \\
$^3$  Leiden Observatory, P.O. Box 9513, 2300 RA, Leiden, The Netherlands \\
$^4$ Astrophysics Group, Imperial College, Blackett Laboratory,
            Prince Consort Road, London, SW7 2BZ \\
$^5$Service d'Astrophysique, Saclay, 91191, Gif-sur-Yvette, Cedex,
France \\ 
$^6$SISSA, Via Beirut 2-4, Trieste, Italy \\
$^7$Osservatorio Astronomico de Padova, Vicolo dell'Osservatorio 5,
I-35 122, Padova, Italy \\
$^8$Max-Planck-Institut f\"ur Extraterrestrische Physik,
Giessenbachstrasse, D-8046, Garching bei Munchen, Germany \\
$^9$Institute for Astronomy, University of Edinburgh, Blackford Hill,
Edinburgh, EH9 3HJ \\
$^{10}$ Max Plank Institute, Heidelberg, Germany \\
$^{11}$ Institut d'Astrophysique Spatiale, Bat. 121, Universit\'e Paris XI,
F-91405 Orsay Cedex, France \\
$^{12}$ Institut d'Astrophysique de Paris, CNRS, 98 bis Bd. Arago,
F-75014, Paris, France\\
email: ciliegi,rgm@ast.cam.ac.uk 
}  
       
\date{Accepted ;
      Received ;
      This version 1998 May 28}

\pagerange{\pageref{firstpage}--\pageref{lastpage}}
\pubyear{1998}

\begin{document}

\maketitle

\label{firstpage}

\begin{abstract}

  We have used the Very Large Array(VLA) in C configuration to carry
  out a sensitive 20cm radio survey of regions of sky that have been
  surveyed in the Far Infra-Red over the wavelength range 5-200 microns
  with ISO as part of the European Large Area ISO Survey(ELAIS).  As
  usual in surveys based on a relatively small number of overlapping
  VLA pointings the flux limit varies over the area surveyed. The
  survey has a flux limit that varies from a 5$\sigma$ limit 
  of 0.135mJy over an area 
  of 0.12deg$^2$ to a 5$\sigma$ limit of 1.15mJy or better over the whole
  region covered of 4.22 deg$^2$. In this paper we present
  the radio catalogue of 867 sources. These regions of sky have
  previously been surveyed to shallow flux limits at 20cm with the
  VLA as part of the VLA D configuration NVSS(FWHM=45 arcsec) and
  VLA B configuration FIRST(FWHM=5 arcsec) surveys. Our whole survey
  has a nominal 5 sigma flux limit a factor of 2 below that of
  the NVSS; 3.4 deg$^2$ of the survey reaches the  nominal flux 
  limit of the FIRST
  survey and 1.5 deg$^2$ reaches to 0.25 mJy, a factor of 3 below the nominal
  FIRST survey limit. In addition our survey is at
  resolution intermediate between the two surveys and thus
  is well suited for a comparison of the reliability and
  resolution dependent surface brightness effects that 
  affect interferometric radio surveys.
  We have carried out a 
  a detailed comparison of the reliability of our own survey  
  and these two independent surveys in order to assess the 
  reliability and completeness of each survey. 

% These comparisons
%  between our maps and those of FIRST and NVSS will be generally useful.

\end{abstract}

\begin{keywords}
radio continuum: general - galaxies
\end{keywords}

\section{Introduction}

The Infrared Space Observatory (ISO, Kessler et al. 1996), 
launched in November 1995 was
the successor of the Infrared Astronomical Satellite (IRAS) 
and provided unparallel sensitivity in mid to
far infrared wavelengths ($i.e.$ 5--200 $\mu m$).  The European Large-Area
ISO Survey (ELAIS, Oliver et al. 1997, Oliver et al. 1998 in preparation) 
is a project that used ISO to carry out a deep wide angle survey 
at wavelengths of 6.7, 15, 90 and 175 $\mu m$. The 6.7 and 15 $\mu m$ surveys 
were carried out with the
ISO-CAM camera (Cesarsky et al. 1996) with the aim to reach a 
5$\sigma$ sensitivity of $\sim$2mJy at 15microns. 
The $90\mu m$ and 175$\mu m$ surveys used the
ISO-PHOT camera (Lemke et al. 1994)  with the aim to reach a 5$\sigma$
sensitivity of $\sim$25mJy.
At these limits, we expect ISO to be confusion limited 
at 90 $\mu m$ and 175 $\mu m$ by galaxies and
galactic cirrus emission and hence this survey should be the deepest FIR
survey possible with the satellite.  

The area covered in the ELAIS survey is
$\sim 13$ square degrees at 15 and 90 microns, $\sim 7$ square degrees
at 6.7 microns and $\sim 3$ square degrees at 175microns. 

The ELAIS survey is $\sim$50 times deeper at 
5-20$\mu m$ than IRAS. Thus our survey will allow us to explore
IRAS-like populations to higher redshift and possibly unveil
new classes of objects or unexpected phenomena. 
We expect to detect thousands of galaxies, many of 
which will be at high redshifts and undergoing vigorous star formation. 
The expected large number of high-z IR galaxies 
should provide vital information about the star formation rate
out to z=1 and possibly earlier.

The spatial resolution of ISO will be insufficient to properly
identify optically faint objects. At 15 microns, the survey resolution is
$\sim$ 10 arcsec and at 90 microns it will be about one arc minute.
Complementary radio data will play a crucial role in identifying many of
the most interesting objects, as they did in the early days of X-ray
astronomy (e.g. Cyg X-1) and in more recent times for IRAS (e.g.
IRAS F10214+4714 (Rowan-Robinson et al. 1991).  

In this paper we report the description of the radio observations 
obtained in the three ISO-ELAIS survey regions in the northern
celestial hemisphere (N1\_1610+5430, N2\_1636+4115 and N3\_1429+3306). 
The observations are made with the Very Large Array (VLA) radio telescope
at 1.4GHz (20cm) in the VLA C-configuration (maximum baseline 11km)
with a resolution (FWHM) of $\sim15$ arcsec.  The aim of these VLA
observation was to obtain an uniform covering of the ELAIS regions with
a rms noise limit of $\sim$50 $\mu$Jy. 
These VLA
observations will be  essential in the optical identification phase
of the ELAIS sources and in assessing the reliability of the
ELAIS source lists.

Moreover, with a radio survey it will be possible to investigate 
the radio/far--infrared correlation in star forming galaxies to  flux
densities deeper than those reached by IRAS. Helou, Sofier \& Rowan-Robinson
(1985) noted a strong correlation between radio and far infrared flux 
for star forming galaxies, valid over a very wide range of infrared 
luminosities, and this has been confirmed in many other studies
(e.g. Wunderlich, Klein \& Wielebinski 1987; Condon, Anderson \& Helou
1991). The radio emission is interpreted as a synchrotron radiation 
from relativistic electron which have leaked out of supernova remnants.
It is expected that this correlation should extend below the IRAS flux 
level since the majority of the sub-mJy radio sources have been identified 
with faint blue galaxies whith spectra similar to those of star forming 
objects (Benn et al. 1993). 

In addition, combining deep radio and optical data with the ISO survey
fluxes will provide information on the trivariate IR-radio-optical
luminosity function and its evolution and the contribution of
starburst galaxies to the sub-mJy radio source counts.  The ratio of
the FIR emission and radio emission will also allow is to investigate
the physical origin and spatial distribution of the energy sources in
the detected objects in the same way that VLA maps have been central
to our understanding of the origin of IRAS sources.

Finally, this survey, due to its depth and extension, is very 
important also as radio survey in its own right. In fact, the selected sample 
is large and deep enough to constitute a statistically significant 
sample of sub-mJy radio sources, whose nature and characteristics 
are still a major topic in observational cosmology (see Windhorst, 
Mathis \& Neuschaefer 1990, Fomalont et al. 1991, Rowan-Robinson 
et al. 1993, Gruppioni et al. 1997)

\section{Radio observations}

\subsection{Choice of observing frequency and VLA configuration}

The VLA C--configuration and the observing frequency of 1.4 GHz 
give the optimum resolution to acquire the kind
of radio data that we need.  Whilst less prone to surface brightness
effects , the VLA D configuration is confusion-limited at the fluxes
we wish to attain (the 5 $\sigma$ confusion limit in D configuration 
is 0.4 mJy/beam).  With the C configuration and a frequency of 1.4 GHz
the synthesized beam size (Full Width at Half Power, FWHP) is $\sim$15
arcsec.  The well-defined synthesized beam of the VLA should enable us
to pinpoint optical identifications to  1 arcsec,
except for the  asymmetric multi-components  sources.
The frequency of 1.4 GHz was chosen because at this
frequency the FWHP of the VLA primary beam is
31 arcmin. This allow us to cover the ELAIS field with a relative
small number of pointing centers. In fact it is possible to obtain a
mosaic map with nearly uniform sensitivity if the separation is 31 /
$\sqrt{2}$ $\sim$ 22 arcmin.  Moreover, at 1.4 GHz there will be
contributions from both the steep and flat spectrum population of
radio sources.

\begin{figure}
\centerline{\psfig{figure=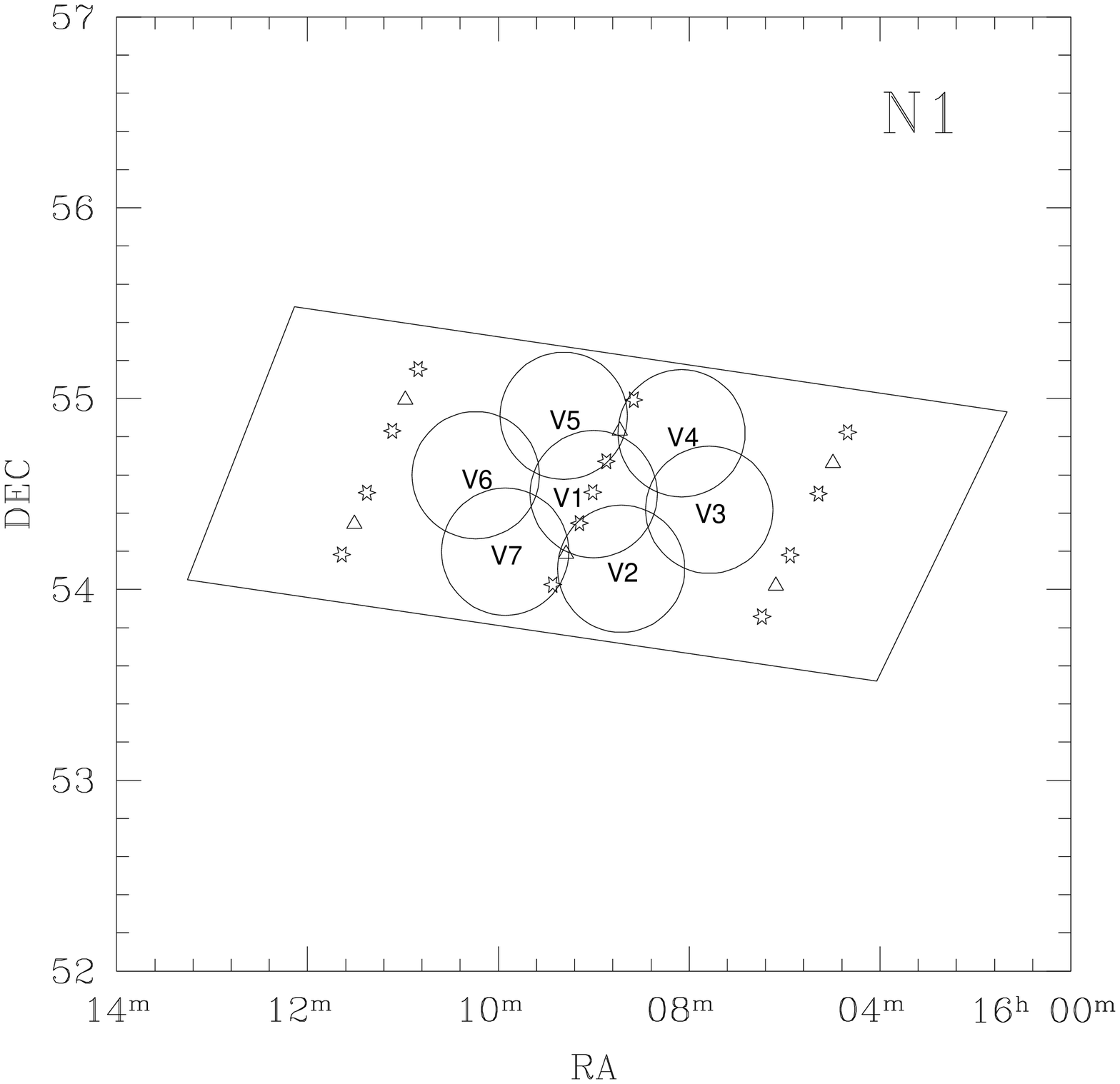,height=7cm} }
\centerline{\psfig{figure=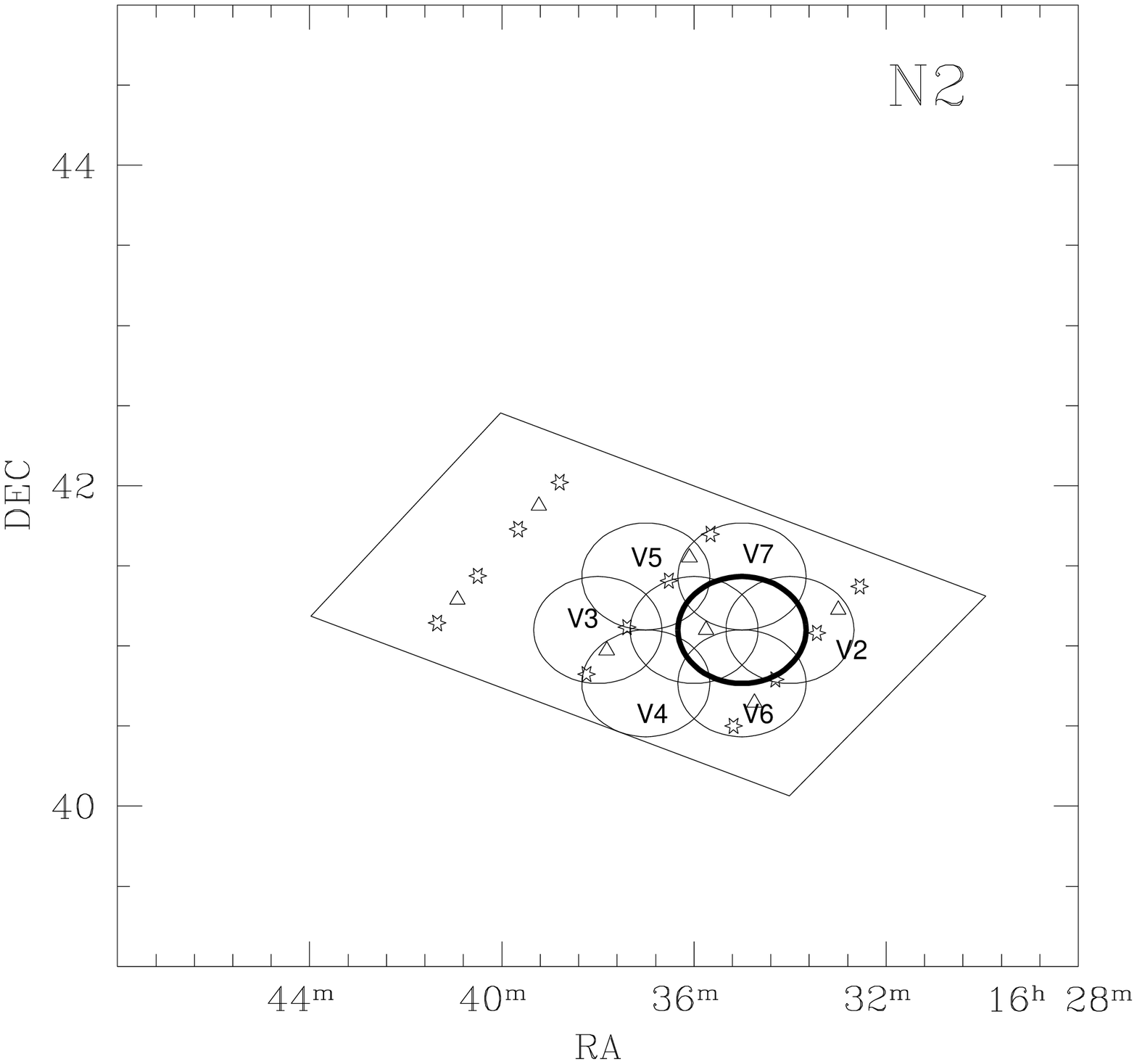,height=7cm} }
\centerline{\psfig{figure=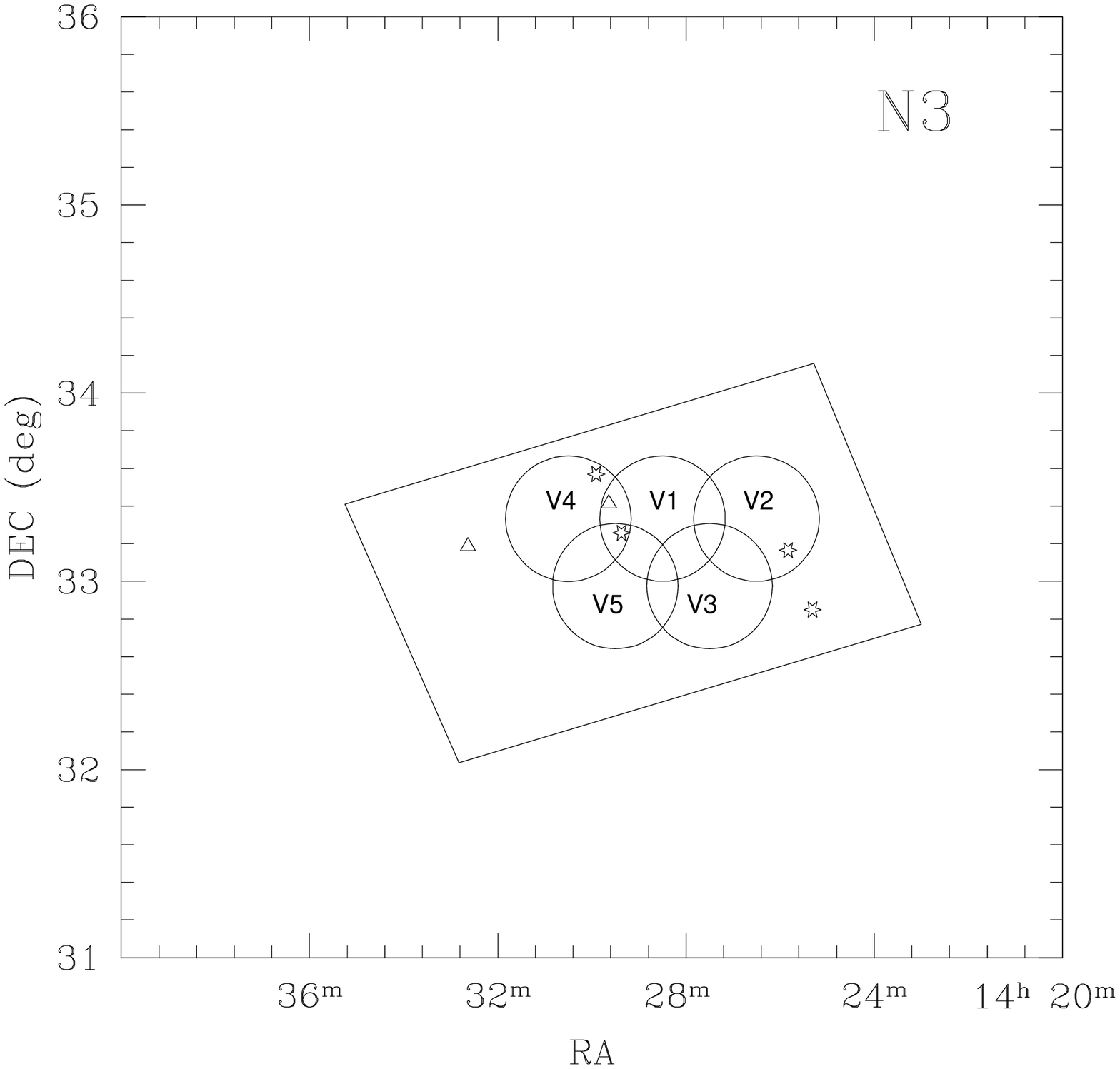,height=7cm} }

\footnotesize
\caption[Layout and status of ELAIS survey fields]
{The sky position and orientation of the ISO
survey regions. Each outer rectangle is 120$^{\prime} \times 80^{\prime}$.
The triangles show the ISO CAM pointings, while the asterisk
show the ISO PHOT pointings.
The circles show the regions we mapped with the VLA.
The circles are drawn with radii R$= 20^{\prime}$, where the VLA 
power sensitivity is $\sim$30\% of the central value. The thick 
circle in N2 is the deep pointing N2 VD (see text for more details).}

\label{fig_elais_grids_status}

\end{figure}

Our observations are made in spectral line mode using two different IF channels
centered at 1.3649 GHz (IF1) and 1.4352 GHz(IF2). Each
IF channel has a bandwidth of 18.75 MHz, which is
subdivided into 7 spectral line
channels evenly spaced in frequency across the bandwidth of the input
IF channel. Therefore, we have a total of 14 spectral line channels (7
for each IF) with a bandwidth of 2.68 MHz each. We decided to use the
line mode in order to facilitate wide field mapping and avoid the
effect of interference.  The total bandwidth available is thus
37.5 MHz, narrower than the 50MHz used in continuum mode.
A narrower bandwidth means a worse sensitivity. In our case
we lose about $\sim$25\% in sensitivity or 15\% in signal
to noise near the pointing centers because the total bandwidth
is 75\% of that in the continuum
mode. However, we considerably reduce the chromatic aberration (bandwidth
smearing), which reduces the area covered by each pointing in
continuum mode. In addition, the line mode is less susceptible to
narrow interference noise spikes, since one only needs to excise the
channel that is affected rather than loosing the whole IF band,
which is the case in continuum mode.
Since our aim is to obtain an uniform sensitivity over
the whole ELAIS fields and not to obtain a single-field deep survey,
we have opted to use the line mode. 
Moreover, the reduced bandwidth
smearing in line mode will give us more accurate angular sizes for our
sources.

\subsection{Observations}

The VLA observation of the ELAIS regions were carried out in
April 96 (10 hours) and in July 1997 (24 hours). A total of 20 pointings
were made in 20cm spectral line mode. Seven of these pointings are in the ELAIS
field N1, eight in the field N2 and five in N3. The integration time 
of each pointing was $\sim$ 1 hour. This allowed us to obtain, in each 
field, a root mean square (rms) noise of $\sim$ 0.05 mJy (a 5 $\sigma$
limit of 0.25 mJy). Moreover, during the July 1997 run, we observed 
two regions (one in N1 and one in N2) with an integration time of 
$\sim$ 3 hours each. In particular, in N1 we re-observed 
(for $\sim$ 3 hours) the pointing V1 (already observed for one 
hour in April 1996) while in N2 the deep pointing (N2 VD) was 
shifted of 10 arc min from the map center to avoid the presence of a strong 
($>$100 mJy) radio source. 
 In the two deep pointings N1 V1 and N2 VD we reached  
an rms noise of  $\sim$ 0.026 mJy (a 5 $\sigma$ limit of 0.13 mJy). 
Finally, in order to study 
the degradation of the image quality as the off-axis of the sources increase, 
a strong calibrator (the source 3C286) was observed at position offset of 
5,10,15,20,25,30 and 35 arcmin in two orthogonal directions (North-South and 
East-West). The result of this test is reported in section ~\ref{offset}.  
 
In Table~\ref{t_vla_observations} we report
the position of the center of each observation, while 
Figure~\ref{fig_elais_grids_status} 
shows
the sky position and orientation of the ISO survey regions. The
circles show the VLA regions mapped. The circles are drawn
with a radius (R$\sim 20^{\prime}$) where the VLA power sensitivity
is $\sim$30\% of the central value. 

\begin{table*}
 \centering
  \caption{VLA Observations of ELAIS regions}
   \label{t_vla_observations}

\begin{tabular}{cccccccc} 
\hline
Region & Pointing & Observing &  RA    &    DEC & Integration & Theoretical& Observed   \\
       &          & Data      &(J2000) & (J2000)& Time (min)  & rms (mJy)  & rms (mJy)  \\ \hline
& & & & \\
N1     &  V1      & April 96 & 16 10 00.00 & +54 30 00.00  & ~58.4 & 0.052 & 0.049 \\
N1     &  V1      & July~ 97 & 16 10 00.00 & +54 30 00.00  & 175.4 & 0.030 & 0.030 \\
N1     &  V2      & July~ 97 & 16 09 25.51 & +54 06 32.42  & ~57.8 & 0.052 & 0.050 \\     
N1     &  V3      & July~ 97 & 16 07 34.62 & +54 25 05.31  & ~58.5 & 0.052 & 0.051 \\     
N1     &  V4      & July~ 97 & 16 08 09.65 & +54 49 08.14  & ~59.5 & 0.052 & 0.052 \\     
N1     &  V5      & July~ 97 & 16 10 37.76 & +54 54 36.12  & ~59.0 & 0.052 & 0.049 \\     
N1     &  V6      & July~ 97 & 16 12 28.43 & +54 35 55.56  & ~59.0 & 0.052 & 0.051 \\     
N1     &  V7      & July~ 97 & 16 11 51.23 & +54 11 54.45  & ~57.8 & 0.052 & 0.052 \\[3mm]
N2     &  V1      & April 96 & 16 36 00.00 & +41 06 00.00  & ~65.3 & 0.050 & 0.050 \\
N2     &  V2      & April 96 & 16 34 00.00 & +41 06 00.00  & ~56.8 & 0.053 & 0.052 \\
N2     &  V3      & April 96 & 16 38 00.00 & +41 06 00.00  & ~59.3 & 0.052 & 0.050 \\
N2     &  V4      & July~ 97 & 16 37 00.00 & +40 46 00.00  & ~59.2 & 0.052 & 0.052 \\     
N2     &  V5      & July~ 97 & 16 37 00.00 & +41 26 00.00  & ~59.0 & 0.052 & 0.052 \\     
N2     &  V6      & July~ 97 & 16 35 00.00 & +40 46 00.00  & ~59.5 & 0.052 & 0.053 \\     
N2     &  V7      & July~ 97 & 16 35 00.00 & +41 26 00.00  & ~59.5 & 0.052 & 0.051 \\
N2     &  VD      & July~ 97 & 16 35 00.00 & +41 06 00.00  & 178.4 & 0.030 & 0.050 \\[3mm]
N3     &  V1      & April 96 & 14 28 30.00 & +33 20 00.00  & ~56.5 & 0.053 & 0.053 \\
N3     &  V2      & April 96 & 14 26 30.00 & +33 20 00.00  & ~58.7 & 0.052 & 0.053 \\
N3     &  V3      & July~ 97 & 14 27 30.00 & +32 58 30.00  & ~58.9 & 0.052 & 0.052 \\     
N3     &  V4      & July~ 97 & 14 30 30.00 & +33 20 00.00  & ~59.5 & 0.052 & 0.050 \\     
N3     &  V5      & July~ 97 & 14 29 30.00 & +32 58 30.00  & ~58.0 & 0.052 & 0.051 \\ 
\hline
\end{tabular}     
 
\end{table*}

\section{Data Reduction}

All the data were analyzed with the NRAO {\tt AIPS} reduction
package.  The data were calibrated using 3C286 as primary flux density
calibrator (assuming a flux of 15.04 Jy at 1.3649 GHz (IF1) and
14.70 Jy at 1.4352 GHz (IF2)) and the sources 1549+506 and 1635+381
as secondary calibrators. The task {\tt UVFLAG} was used to ``flag''
(delete) the corrupted data ($e.g.$ bad integration, non operating antennas,
high amplitudes due to interferences $etc$) while the tasks {\tt
VLACALIB} and {\tt GETJY} were used to calibrate the data and to
determine the source flux densities.  Finally each observation was
cleaned using the task {\tt IMAGR}.

\subsection{Root mean square (rms) map noise of the single pointings}

\begin{figure}
\centerline{\psfig{figure=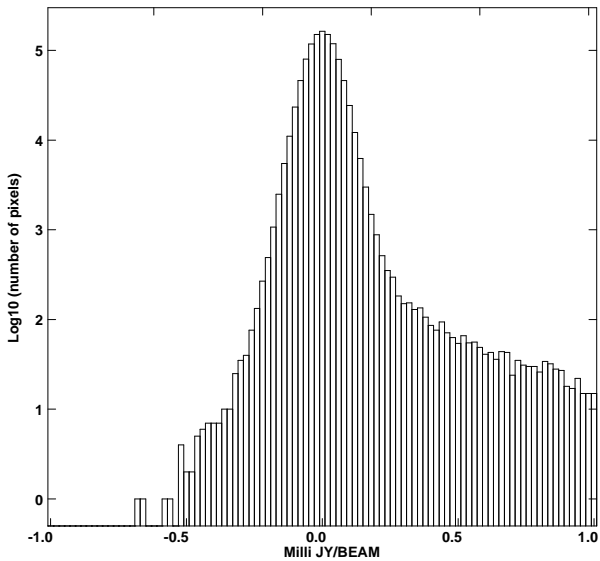,width=8cm} }
\caption
[Histogram of VLA map peak flux densities in the pointing N2 V1]
{Logarithmic histogram of VLA map peak flux densities in the pointing N2 V1}
\label{fig_hist_vla_map_noise}
\end{figure}

The integration time of each observation after deletion of the 
corrupted data are reported in  Table~\ref{t_vla_observations}.  
The rms noise of each pointing was estimated using the amplitude 
distribution of the pixel values in the cleaned map before correcting 
for primary-beam attenuation. In 
Figure~\ref{fig_hist_vla_map_noise} we show this distribution 
for the pointing N2 V1. This distribution is the sum
of a Gaussian noise core with a positive-going tail caused 
by discrete sources. Note that the y-axis in 
Figure~\ref{fig_hist_vla_map_noise} is logarithmic.

Thus the rms of the noise  distribution alone 
should be nearly equal to the rms of that distribution obtainable 
by reflecting the negative flux portion of the observed 
amplitude distribution about flux=0. In column VII of 
Table~\ref{t_vla_observations}
we report the theoretical rms noise level (at 1$\sigma$) 
as computed directly from the observing parameters 
(integration time, number of antennas, observing frequency, 
bandwidth and number of $IF$ channels), while in column VIII 
we report the 1$\sigma$ rms noise level obtained by fitting 
the noise distribution of each pointing. As shown in 
Table~\ref{t_vla_observations}
there is a very good agreement  between the observed and the 
theoretical rms noise levels. 

\subsection{Mosaic maps}

\begin{figure}
%\begin{minipage}{160mm}
\centerline{\psfig{figure=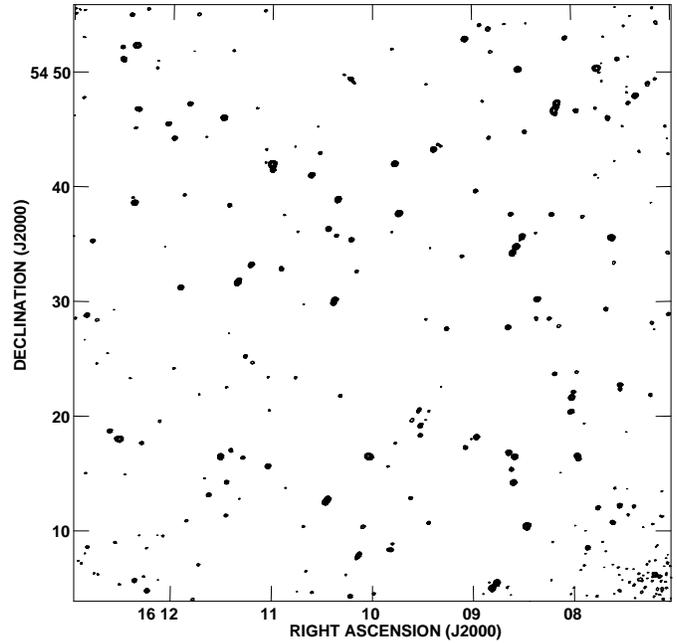,width=9cm} }
%\end{minipage}
%\begin{minipage}{160mm}
%\centerline{\psfig{figure=N2_26.PS,width=8cm} }
%\end{figure}
%\begin{figure}
%\end{minipage}
%\begin{minipage}{160mm}
%\centerline{\psfig{figure=N2DEEP.PS,width=8cm} }
%\end{minipage}
%\begin{minipage}{160mm}
%\centerline{\psfig{figure=N3_26.PS,width=8cm} }
%\end{minipage}

\caption{Central 26$\times$26 arcmin$^2$ = 520$\times$520 pixels of the 
mosaic map N1. The contour level are 0.25,0.50,0.75,1.00,
1.25 ... mJy. The noise at the corners of the image (see for example 
the low right corner) is due to the increase of the rms noise as  the
off-axis value increase}

\label{contour_map}

\end{figure}

Using the {\tt AIPS}  task {\tt LGEOM}, {\tt HGEOM} and {\tt LTESS}
we have combined all the observations and we have created a mosaic map 
for each field (N1, N2 and N3). A special procedure has been adopted 
for N2. The presence of a deep observation not in the center of 
the radio map (see above and Figure~\ref{fig_elais_grids_status}) 
makes the noise of this map strongly irregular. To simplify the 
extraction of the sources we have created two different mosaic maps. 
In the first map we have combined all the observation in N2 
excluding the deep pointing, while in the second one we have combined 
the deep pointing N2 VD with all the surrounding pointings 
(N2 V1, N2 V2, N2 V6 and N2 V7, see Figure~\ref{fig_elais_grids_status}). 
In this way we have obtained two mosaic maps with a regular noise: lower 
in the map center and higher in the outer regions. 

An example of our map is given in  Figure~\ref{contour_map} where we 
show the  contour map of the central region 
(26$\times$26 arcmin$^2$) of the  mosaic map N1.

\subsection{Noise of the mosaic maps}

\begin{figure*}
\centerline{\psfig{figure=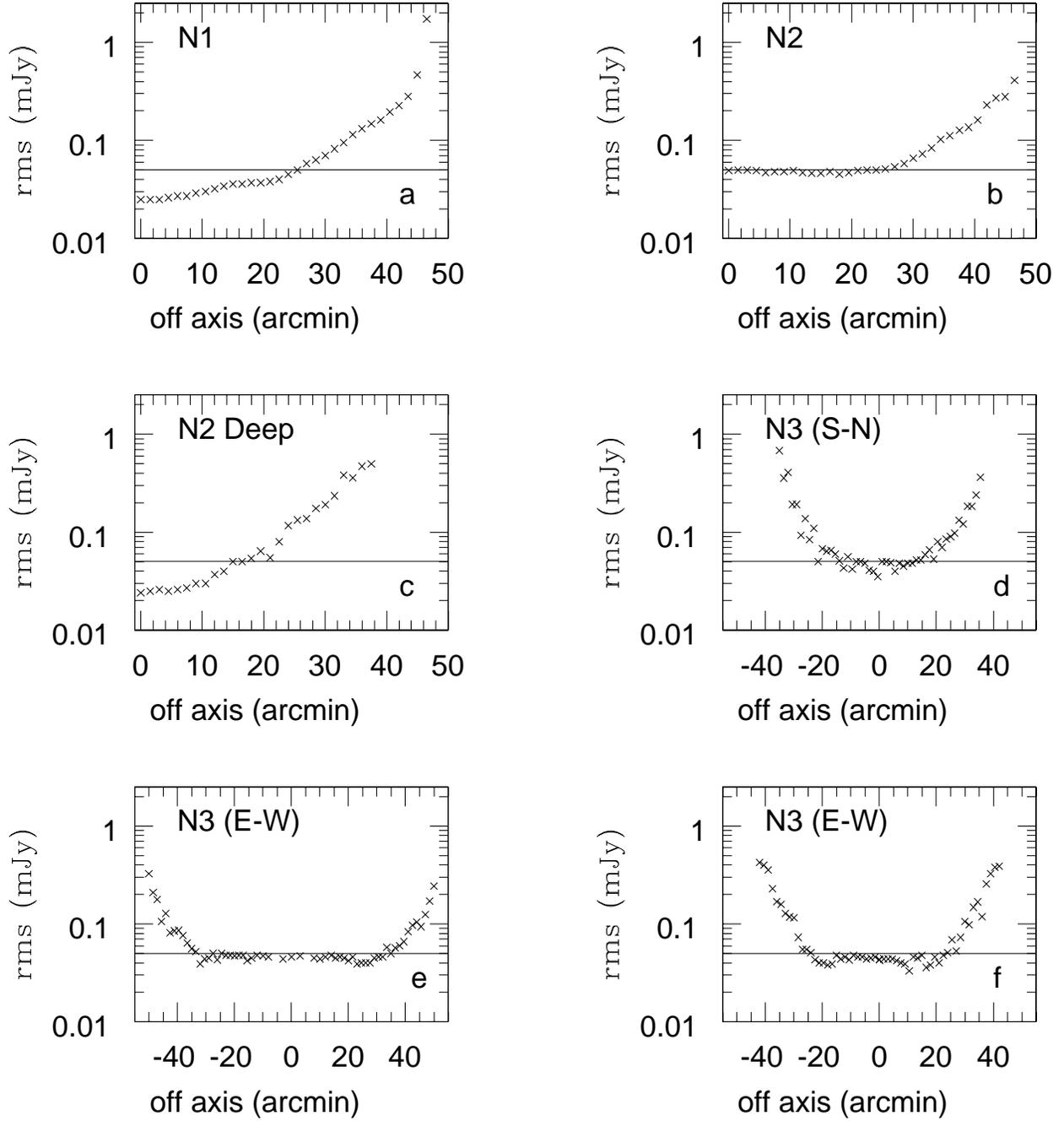}}
\caption{The rms noise value as function of the off-axis value. 
The solid line represent an rms value of 0.05 mJy. See text for more details.}
\label{rms_off}
\end{figure*}

We analyzed the noise properties
in each mosaic maps (N1, N2, N2 Deep and N3).  
As expected, N1, N2 and N2 Deep have a regular noise
distribution: a circular central region with a flat 
noise surrounded by concentric annular region, where the noise 
increases for increasing distance from the center (the off-axis 
value). No structures or irregularities were found in the rms maps. 
In Figure~\ref{rms_off} we plot the variation of the 
rms as function of the off-axis value for N1, N2 and N2 Deep. 
Due to the different number of pointings in N3 (five pointings 
instead of seven, see Figure~\ref{fig_elais_grids_status})
the mosaic map in this field has a different shape and the noise distribution 
in the map has a semi-circular shape instead of a circular 
one. In Figure~\ref{rms_off}d,e,f we plot the variation of the 
rms value in three different slices. In Figure~\ref{rms_off}d we 
plot the rms in the North-South direction, in 
Figure~\ref{rms_off}e the rms in a slice that connects the 
center of the three northern pointings (N3 V1, N3 V2 and N3 V4), 
while in Figure~\ref{rms_off}f we report the rms value in a 
slice that connects the center of the two souther pointings 
(N3 V3 and N3 V5). As for N1 and N2, no structures or irregularities
were found in the rms map of N3.

\section{Source Catalogue}

\subsection{Regions for the source extraction}

\begin{figure*}
\centerline{\psfig{figure=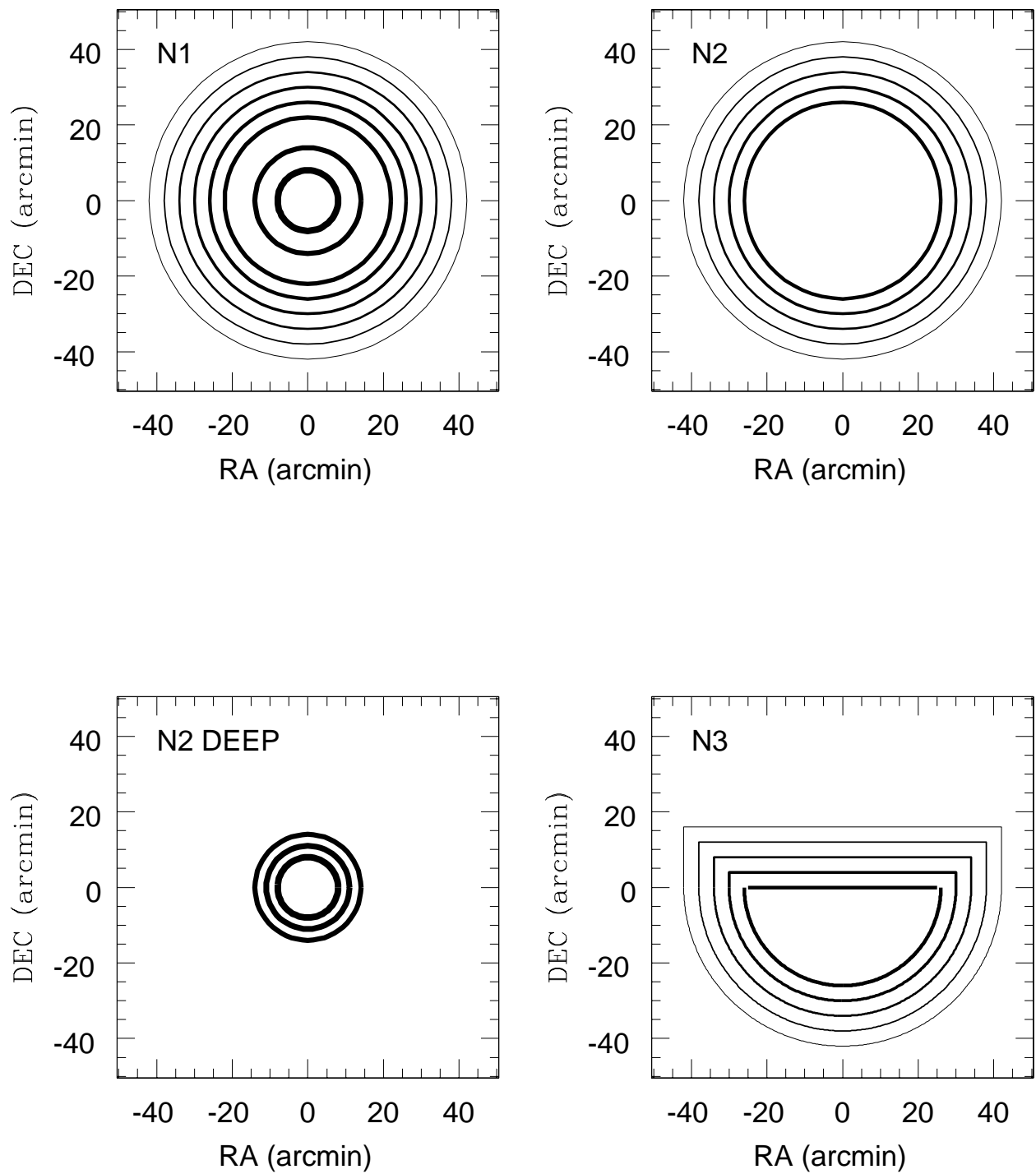}}
\caption{The region used for the source extraction. The lines 
with the same thickness enclose regions with the same rms 
noise value (see Table~\ref{zone_tab}) }
\label{rms_zone}
\end{figure*}

Using Figure~\ref{rms_off} we determined the regions with 
constant rms noise to be used for the source extraction.
Due to the circular shape of the rms map, we searched for 
radio sources in a circular central region plus concentric 
annular regions to an off-axis value of 42 arcmin. 
Outside this area, the beam attenuation significantly 
increases the limiting flux (see Figure~\ref{rms_off}). 
In each region the rms used for source extraction is the 
highest rms in the region. In this way we are sure that 
all the sources extracted are above the fixed threshold in 
the region (for example 5 times the rms value). 
The sizes of these regions are reported in Table~\ref{zone_tab}
and in Figure~\ref{rms_zone}.

\begin{figure}
\centerline{\psfig{figure=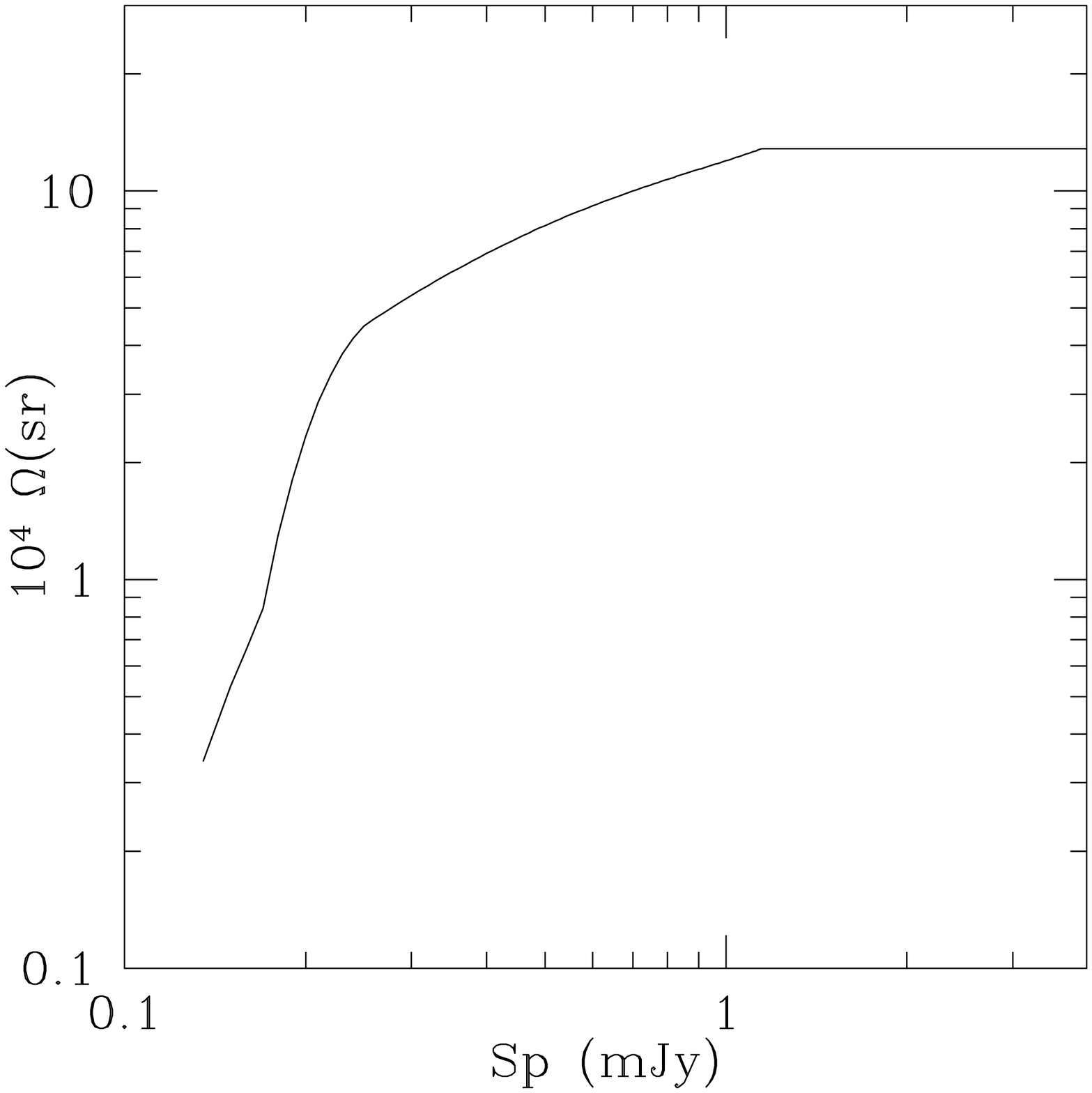,width=8cm}}
\caption{Solid angle over which a source with detected peak
flux density S$_p$ can be counted.}
\label{area}
\end{figure}

Using the data reported in Table~\ref{zone_tab} we obtained the 
solid angle $\Omega(S_p)$ as function of the peak flux density
$S_p$ covered by our survey. This effective survey area is 
shown in Figure~\ref{area}. 

\vspace{5mm}

\begin{table}
\caption[The regions used for the sources extraction]
{The regions used for the sources extraction}
\label{zone_tab}

\begin{center}

\begin{tabular}{lcccccc} 
& & & & \\ \hline
Region & R$_{inn}$ & R$_{out}$  & Size    & Total  & rms  & 5 $\sigma$ limit \\
       & ($\prime$)& ($\prime$) & deg$^2$ &deg$^2$ & mJy  & mJy \\ \hline
{\bf N1} &   & \\
1      &    0     &      8      & 0.0559  & 0.0559  &  0.027 & 0.135 \\
2      &    8     &     14      & 0.1151  & 0.1710  &  0.034 & 0.170 \\
3      &    14    &     22      & 0.2513  & 0.4223  &  0.038 & 0.190 \\
4      &    22    &     26      & 0.1677  & 0.5900  &  0.050 & 0.250 \\
5      &    26    &     30      & 0.1955  & 0.7855  &  0.070 & 0.350 \\ 
6      &    30    &     34      & 0.2234  & 1.0089  &  0.100 & 0.500 \\ 
7      &    34    &     38      & 0.2513  & 1.2602  &  0.150 & 0.750 \\ 
8      &    38    &     42      & 0.2793  & 1.5395  &  0.230 & 1.150 \\[3mm]
{\bf N2} &   & \\                                             
1      &    0      &    26      & 0.5900  & 0.5900  &  0.050 & 0.250 \\
2      &    26     &    30      & 0.1955  & 0.7855  &  0.070 & 0.350 \\ 
3      &    30     &    34      & 0.2234  & 1.0089  &  0.100 & 0.500 \\ 
4      &    34     &    38      & 0.2513  & 1.2602  &  0.150 & 0.750 \\ 
5      &    38     &    42      & 0.2793  & 1.5395  &  0.230 & 1.150 \\[3mm]
{\bf N2 D} &   & \\                                           
1      &    0      &     8      & 0.0559  & 0.0559  &  0.027 & 0.135 \\
2      &    8      &    11      & 0.0497  & 0.1056  &  0.030 & 0.150 \\
3      &    11     &    14      & 0.0655  & 0.1711  &  0.038 & 0.190 \\
{\bf N3} &   & \\                                             
1      &    0      &    26      & 0.2950  & 0.2950  &  0.050 & 0.250 \\
2      &    26     &    30      & 0.1645  & 0.4595  &  0.070 & 0.350 \\ 
3      &    30     &    34      & 0.1961  & 0.6556  &  0.100 & 0.500 \\ 
4      &    34     &    38      & 0.2279  & 0.8835  &  0.150 & 0.750 \\ 
5      &    38     &    42      & 0.2597  & 1.1432  &  0.230 & 1.150 \\[3mm]
\hline
\end{tabular}
 
\end{center}

\end{table}

\subsection{Source detections}

Within each region we searched for radio sources up to a peak flux
density $\geq$ 5 times the rms value of the region. The sources were
extracted using the task {\tt SAD} (Search And Destroy) which attempts
to find all the sources whose peaks are brighter than a given level.
For each selected source the flux, the position and the size are
estimated using a least square Gaussian fit.  However, for faint
sources a 
Gaussian fit may be unreliable (see Condon 1997, for an
extensive discussion about errors in Gaussian fits). For this reason
we decided first to run the task {\tt SAD} with a flux limit of 3 times 
the local rms value (see Table ~\ref{zone_tab}). Subsequently, we derived
the peak flux of faint sources (those
detected with $3\sigma < S_{peak} < 7\sigma$) using a second degree
interpolation (task {\tt MAXFIT}). Only the sources with a {\tt MAXFIT} 
peak flux density  $\geq$ 5 $\sigma$ were included in the sample. 
For these faint sources the total flux density was obtained using the 
task {\tt IMEAN}, which integrates the map value in a specific
rectangle, while for all the other parameters (major axis, minor axis and 
position angle) we used the values obtained with the Gaussian fit. 
For irregular resolved sources the total flux density
was calculated using the task {\tt TVSTAT} which allows us to use an 
irregular area to integrate the map value.

\subsection{Multiple sources}

Detected sources separated by less than two times the value of our
synthesized beam size ($i.e.$ 25 arcsec) and with a flux ratio lower
than 2 have been considered as an unique source. We adopted this criterion
because the component flux density ratio of physically doubles is 
usually small ($\lsimeq$2) while the projection pairs can have arbitrarily
large flux density ratio (Condon, Condon and Hazard, 1982).

\section{The Source Catalogue}

Considering all the available observations we detected a total of 
867 sources at $\geq$ 5 $\sigma$ level (44 of which have multiple 
components) over a total area of 4.222 deg$^2$. The catalogue with all 
the 867 sources (921 components) reports the name of the source, the 
peak flux density S$_P$ in mJy, the total flux density S$_I$ in mJy, 
the RA and DEC (J2000), the full width half maximum (FWHM) of the 
major and minor axes $\theta_M$ and $\theta_m$ (in arcsec), the positional 
angle PA of the major axis (in degrees) and the off-axis values in the 
VLA map (in arcmin).
The different components of multiple sources are labeled ``A'', ``B'', 
etc., followed by a line labeled ``T'' in which flux and position 
for the total sources are given. For these total sources the 
position have been computed as the 
flux-weighted average position for all the components. 
Table 3 shows the first 
page of the catalogue as an example. 

\begin{figure}
\centerline{\psfig{figure=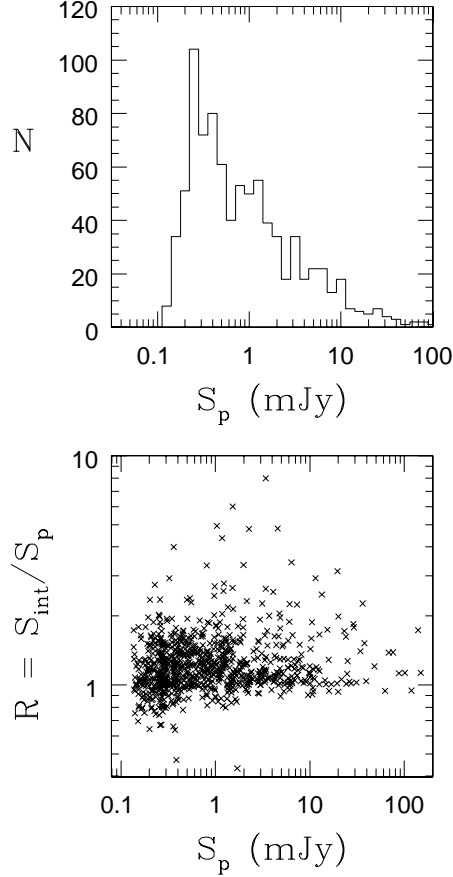,width=14cm}}
\vspace{-2cm}
\caption[Flux distributions and total to peak flux ratio as function 
of peak flux]{Peak flux distributions (upper) and total to peak 
flux ratio as function of peak flux (lower) for all the     
radio sources detected in N1, N2 and N3}
\label{flux_ratio}
\end{figure}

\begin{figure}
%\vspace{22cm}
%\begin{minipage}{160mm}
\centerline{\psfig{figure=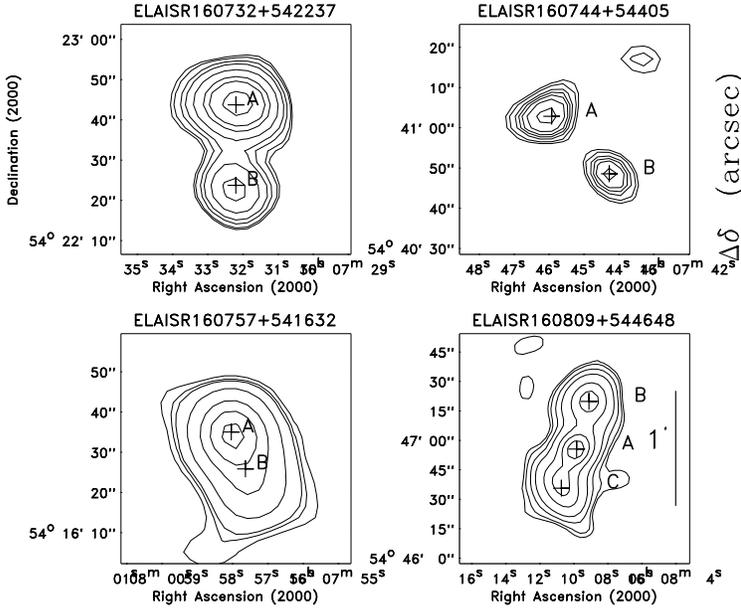,width=9cm}}
\caption{ SMALL EXAMPLE (This figure is available 
in its entirety in http://www.ast.cam.ac.uk/$\sim$ciliegi/elais/papers/)
 : Contour maps for multiple 
sources in N1. Contour levels 
are 5,6,7,8,10 .... times the local rms noise value (see Table~\ref{zone_tab})
Each map is $1^{\prime} \times 1^{\prime}$  except  for very extended sources 
for which the map size is variable. For these sources the vertical 
bar in the map shows 1 arcmin.}
\label{double_contour1}
%\end{minipage}
\end{figure}

In Table~\ref{sources_tab}
we report the number of radio sources detected in N1, N2 and N3, while 
in Figure~\ref{flux_ratio} we show the distribution of the peak flux density
and the total to peak flux ratio as a function of peak flux for 
all the 867 sources. Contour maps of all the 44 double or multiple 
sources are shown in Figure ~\ref{double_contour1}  ~\ref{dN2} and
~\ref{dN3}. 

\begin{figure*}
\vspace{-1.4cm}
\centerline{\psfig{figure=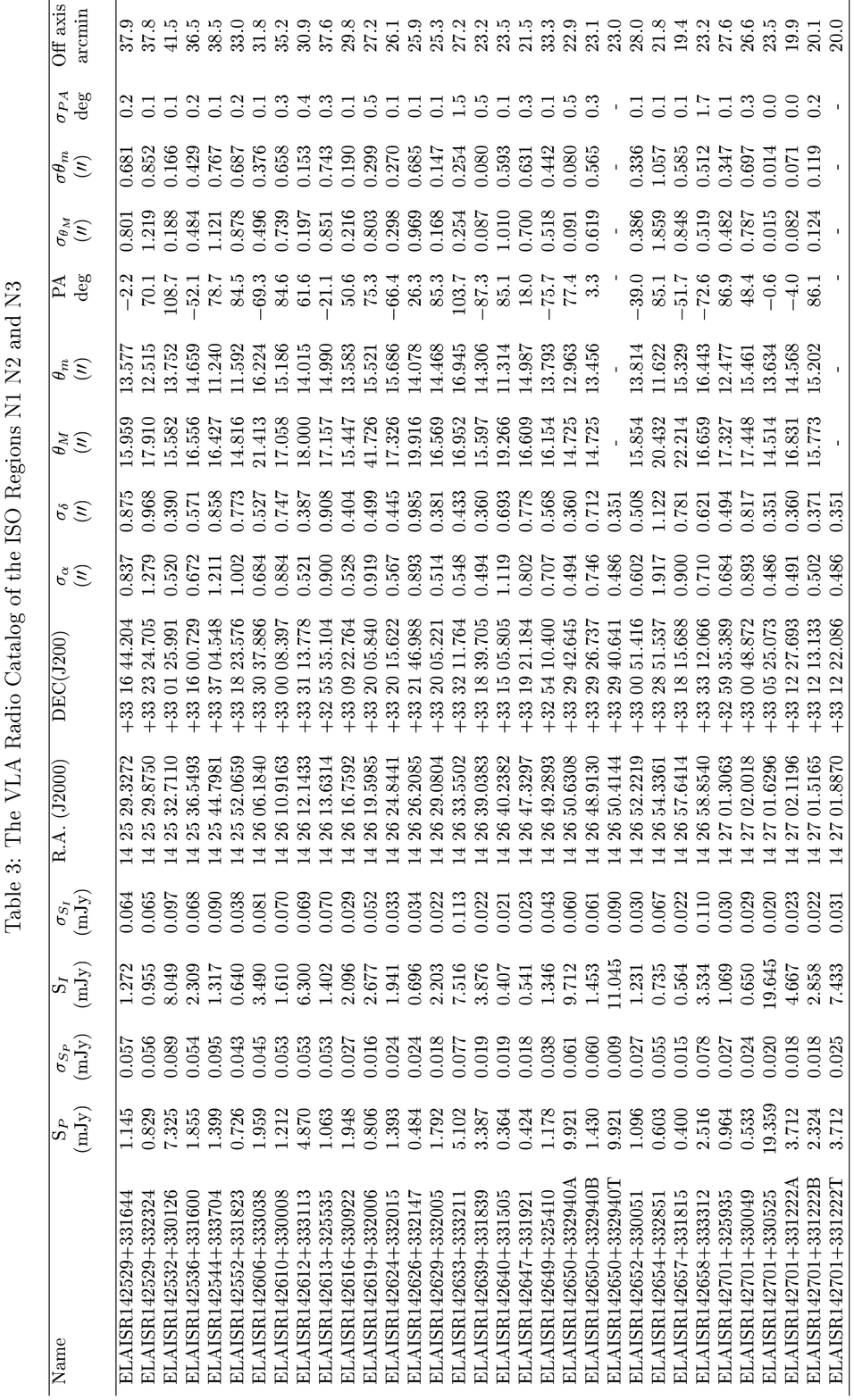,width=18cm}}
%\addtocounter{figure}{-1}
\addtocounter{table}{+1}
\end{figure*}

\begin{figure}
\vspace{1cm}
%\begin{minipage}{90mm}
%\centerline{\psfig{figure=double_cntr_N2.ps}}
\caption{As in Figure \ref{double_contour1} for N2. 
This figure is available 
in http://www.ast.cam.ac.uk/$\sim$ciliegi/elais/papers/}
\label{dN2}
%\end{minipage}
\end{figure}

\begin{figure}
\vspace{1cm}
%\begin{minipage}{90mm}
%\centerline{\psfig{figure=double_cntr_N3.ps}}
%\vspace{-3.5cm}
\caption{As in Figure \ref{double_contour1}  for N3.
This figure is available 
in in http://www.ast.cam.ac.uk/$\sim$ciliegi/elais/papers/}
\label{dN3}
%\end{minipage}
\end{figure}

\begin{table}
\caption[The number of radio sources detected in each field]
{The number of radio sources detected in each field}
\label{sources_tab}

\begin{center}

\begin{tabular}{cccc} 
& & & \\ \hline
Field  & Single   & Multiple & Total \\
       & Sources  & Sources  & Sources \\ \hline
&  & \\
N1 & 345 & 16 & 361 \\
N2 & 305 & 16 & 321 \\
N3 & 173 & 12 & 185 \\
\hline

\end{tabular}
 
\end{center}

\end{table}

\subsection{Sources parameters and their uncertainties}

\subsubsection{Positions and angular sizes}

\begin{figure}
\centerline{\psfig{figure=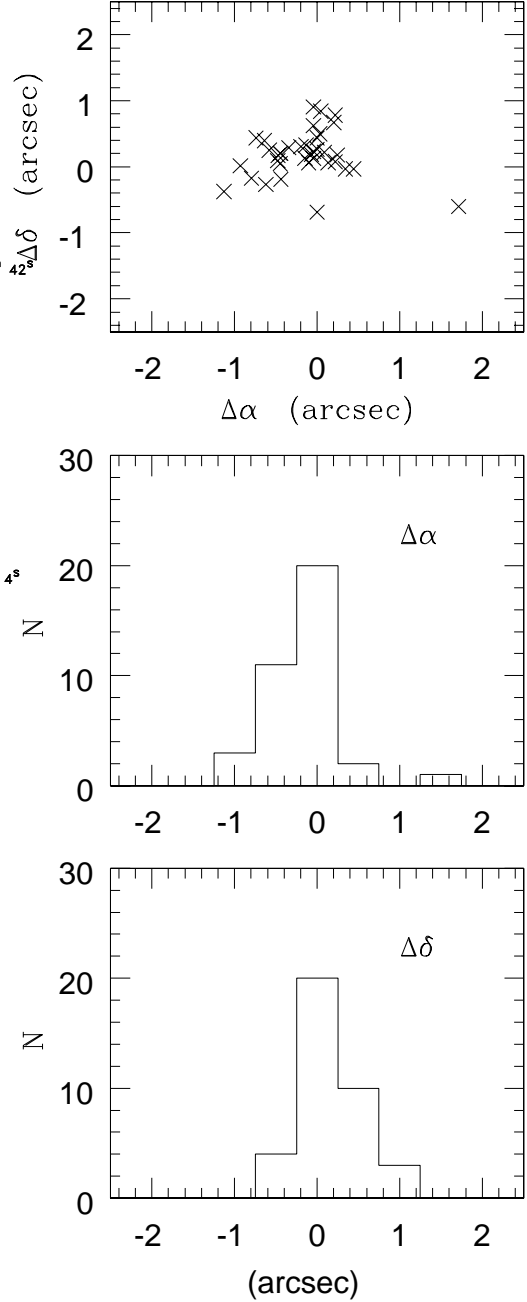,height=18cm}}
\caption{Position errors for strong point sources (37 sources in 
our survey and in the $FIRST$ survey with flux radio greater than 
10 mJy)}
\label{histo_strong}
\end{figure}

Following Condon 1997, the error on the sources parameters reported in the 
catalogue have been calculated using : 
\begin{equation}
\frac{\sigma^2_{S_P}}{S^2_P}=\frac{\sigma^2_{S_I}}{S^2_I}=
\frac{\sigma^2_{\theta_M}}{\theta^2_M}=\frac{\sigma^2_{\theta_m}}{\theta^2_m}=
\frac{\sigma^2_{PA}}{2}\left( \frac{\theta^2_M - \theta^2_m}{\theta^2_M \theta^2_m} \right)^2
=\frac{2}{\rho^2}
\end{equation}
where $\rho$ is the signal to noise ratio given by
\begin{equation}
\rho=\frac{\pi}{8 ln2} \frac{\theta_M \theta_m S_P}{h^2 \sigma^2_{map}}
\end{equation}
Equations 1 and 2 are the master equations for estimating variances of the 
parameters derived from a two dimensional Gaussian fit on an image with a  
noise variance $\sigma^2_{map}$ and pixel size h. In our maps 
$h$=3 arcsec,  while column 5 of Table ~\ref{zone_tab} reports the value of  
$\sigma_{map}$  as function of the off-axis value. 
The rms position error are given by (Condon et al. 1998):
\begin{equation}
\sigma^2_{\alpha} = \varepsilon^2_{\alpha} + \sigma^2_{\theta_M} \sin^2(PA) + 
\sigma^2_{\theta_m} \cos^2(PA)
\end{equation}
\begin{equation}
\sigma^2_{\delta} = \varepsilon^2_{\delta} + \sigma^2_{\theta_M} \cos^2(PA) + 
\sigma^2_{\theta_m} \sin^2(PA)
\end{equation}
The mean image off-set $<\Delta\alpha>$, $<\Delta\delta>$ and rms calibration
uncertainties $\varepsilon^2_{\alpha}$ and $\varepsilon^2_{\delta}$ are best
determined by comparison with  accurate position of sources strong enough 
that the noise plus confusion terms are much smaller than the calibration
terms. We used the 37 single compact sources stronger than 10 mJy 
found in the $FIRST$ (Faint Images of the Radio Sky at Twenty cm) radio 
survey (Becker et al. 1997). Their off-sets $<\Delta\alpha>$ and  
$<\Delta\delta>$ are shown in Figure ~\ref{histo_strong}. The mean 
off-sets of our map are $<\Delta\alpha>=-0.15^{\prime\prime}$ and 
$<\Delta\delta>=0.18^{\prime\prime}$. The sources coordinate reported 
in the catalogue (see Table 3) have been corrected for these off-sets. 

\begin{figure}
\centerline{\psfig{figure=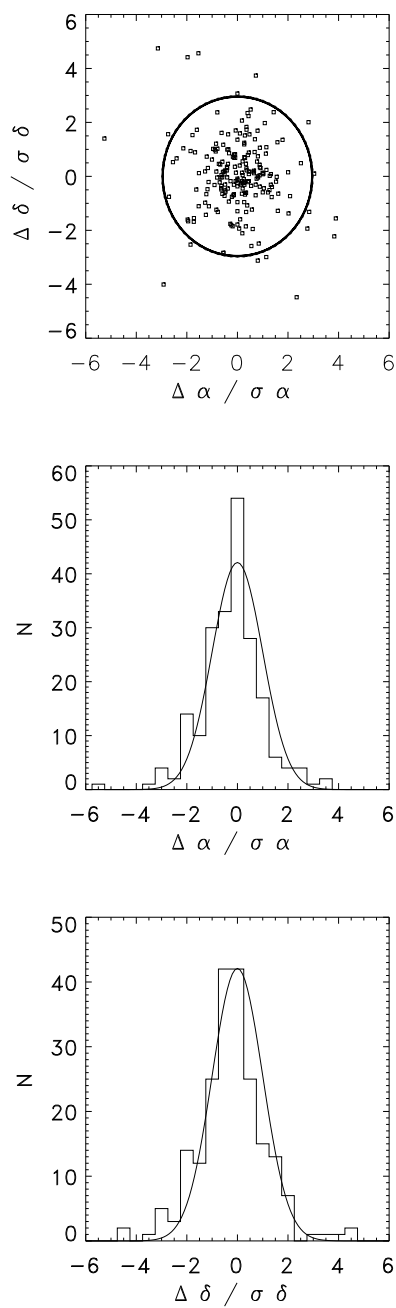,height=18cm}}
\caption{Position errors for the 211 point sources stronger than 1 mJy
in common with the $FIRST$ survey, in unit 
of the calculated position uncertainties $\sigma_{\alpha}$ and 
$\sigma_{\delta}$, along with the 90\% confidence error circle ({\it upper panel}). The smooth curves in the central and lower panel represent
the expected Gaussian of zero mean and unit variance.}
\label{histo_medium}
\end{figure}

From the distribution of $<\Delta\alpha>$ and  $<\Delta\delta>$ shown in 
Figure ~\ref{histo_strong} we have estimated the rms calibration errors: 
$\varepsilon_{\alpha}=0.486^{\prime\prime}$ and 
$\varepsilon_{\delta}=0.351^{\prime\prime}$. Their small values show that
our mosaic maps are not affected by relevant geometric distortions due to the 
approximation  of a finite portion of the spherical sky away from the 
instrumental zenith with a bi-dimensional plane (see Perley 1989, Condon 
et al. 1998). Using the mean off-sets and calibration errors obtained using
the 37 sources stronger than 10 mJy in common with the $FIRST$ survey, 
we have calculated the positional errors of all our sources. 
To verify that our positional errors are realistic also for faint 
sources, we used the 211 single compact sources stronger than 1 mJy  
found to be in common with the $FIRST$ survey. The right ascension and 
declination differences $<\Delta\alpha>$ and  $<\Delta\delta>$  were 
divided by the uncertainties $\sigma_{\alpha}$ and $\sigma_{\delta}$ 
(from Equations 3 and 4). As expected, the normalized error distributions 
(Figure~\ref{histo_medium}) are Gaussian with nearly zero mean and unit 
variance, verifying that our positional uncertainties are accurate also 
for sources down to $\sim$1 mJy. 

\begin{figure}
\centerline{\psfig{figure=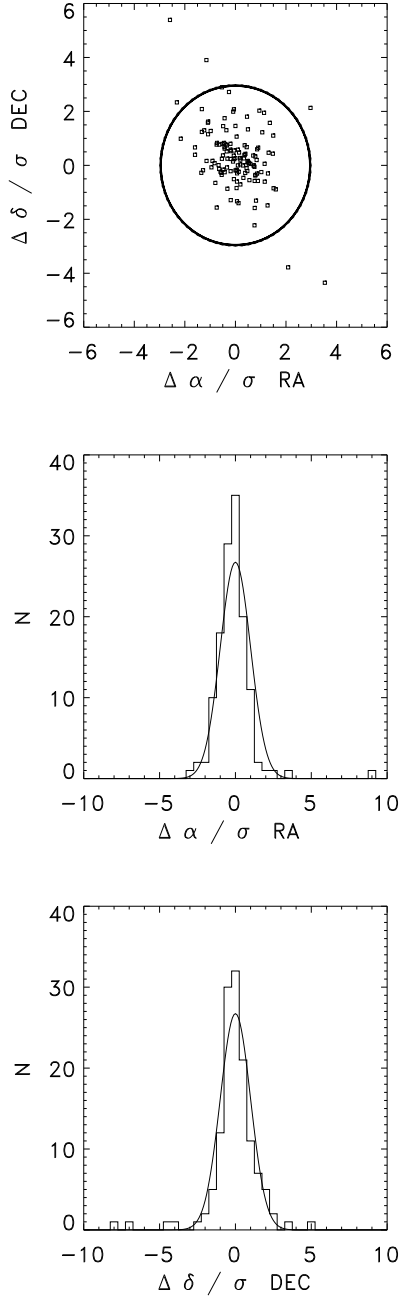,height=18cm}}
\caption{Position errors for the 134 point sources in the 
overlapping region with flux lower  than 1 mJy, in unit 
of the combined position uncertainties $\sigma_{RA}$ and 
$\sigma_{DEC}$, along with the 90\% confidence error circle 
({\it upper panel}). The smooth curves in the central and lower panel represent
the expected Gaussian of zero mean and unit variance.}
\label{histo_weak}
\end{figure}

However, as shown in the flux distribution 
of Figure~\ref{flux_ratio}, the majority of our radio sources have a flux 
lower than 1 mJy. Since we do not have other radio catalogues below 1 mJy 
to repeat out test, we decided to use our observations to test the reliability
of the position errors down to the flux limit of our survey ($\sim$ 0.135 mJy).
In particular, we used the following procedure: (1) Each mosaic map has been 
obtained combining five or more individual pointings, whose centers are 
separated by $\sim$ 22 arcmin. (2) Because in each individual map the FWHM 
of the primary beam is 31 arcmin, there are many overlapping regions where 
sources are detected in two different pointings. (3) We considered the sources
detected in each individual pointing as an independent data set and we used 
the sources in common ($i.e$ the sources detected in the overlapping regions) 
to test the reliability of the positional errors. We have a total of 134 
sources in the overlapping regions with a flux lower than 1 mJy. The right 
ascension and declination differences of these 134 sources were divided by the 
combined errors $\sigma_{RA}$ and $\sigma_{DEC}$, where 
$\sigma^2_{RA}=\sigma^2_{\alpha 1}+\sigma^2_{\alpha 2}$ and 
$\sigma^2_{DEC}=\sigma^2_{\delta 1}+\sigma^2_{\delta 2}$ with 
$\sigma_{\alpha 1}, \sigma_{\alpha 2}, \sigma_{\delta 1}$ and 
$\sigma_{\delta 2}$ from Equations 3 and 4.  As shown in 
Figure~\ref{histo_weak}, also these error distributions for faint sources 
agree with the expected Gaussians (smoothed curves).

\begin{figure}
\centerline{\psfig{figure=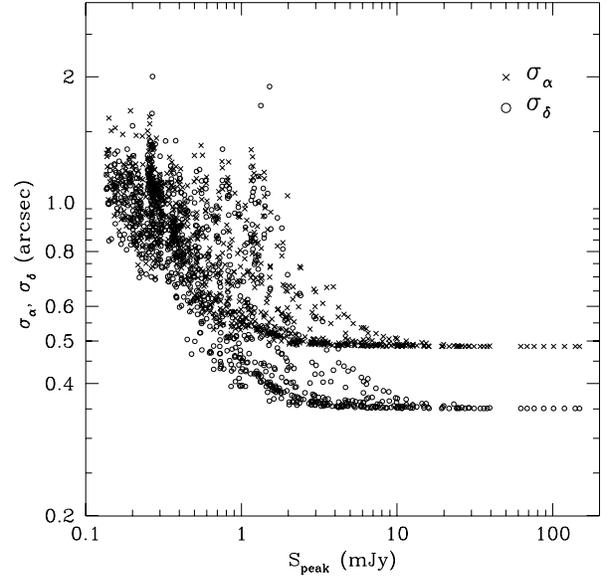,width=8cm}}
\caption{The typical rms position uncertainties 
$\sigma_{\alpha}$ and $\sigma_{\delta}$ for point sources of flux-density
S$_{peak}$.}
\label{ra_dec_error}
\end{figure}

In conclusion the positional uncertainties obtained using Equations 3 and 4 
are accurate also for sources down to the limit of our survey ($\sim$ 0.135
mJy) and they can be used to estimate the reliability of the optical and 
infrared identifications. The typical rms position uncertainties 
$\sigma_{\alpha}$ and $\sigma_{\delta}$ are plotted as function of the 
flux density in Figure ~\ref{ra_dec_error}. 
The positional errors of the 
radio sources are $\sim$ 2 arcsec for the fainter sources
($\sim$0.13 mJy) and $\sim$ 0.6  arcsec for the brighter
sources ($>$ 10 mJy). 

\subsubsection{Bandwidth Smearing}
\label{offset}

\begin{figure}
\centerline{\psfig{figure=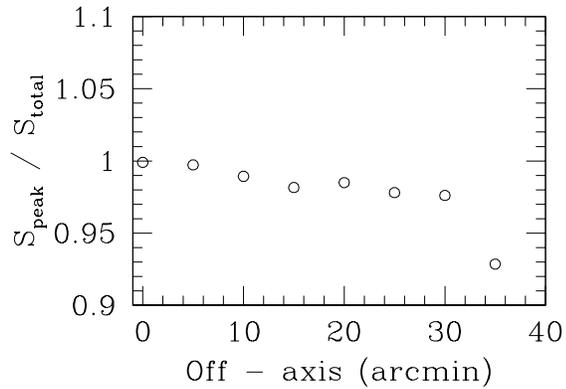,width=9cm}}
\vspace{-2cm}
\caption{Bandwidth smearing reduces the peak flux density S$_P$ of a 
point sources but preserves the integrated flux density S$_I$. 
Abscissa : angular distance from the center (arcmin).
Ordinate : The observed ratio S$_P$/S$_I$ }
\label{smearing}
\end{figure}

The principles upon which synthesis imaging are based are strictly valid 
only for monochromatic radiation. When radiation from a finite bandwidth 
smearing is accepted, aberrations in the image will result. These take the 
form of radial smearing which worsen with increased distance from the 
center. The peak response to a point sources simultaneously declines 
in a way that keeps the integrated flux constant. In other words the 
bandwidth smearing reduces the peak flux density S$_P$ of a source but 
not its integrated flux density S$_I$. However, in our case, since we used 
the spectral line mode where the bandwidth is narrower than in  continuum mode, 
the bandwidth smearing should be negligible.  To verify that this is the case, 
a strong calibrator (the
source 3C286) was observed at position offset of 5,10,15,20,25,30,35 
arcmin in two orthogonal directions (North-South and East-West) and we made 
a Gaussian fit of the source in each of the offset position. 
The ratio between peak and total flux density (S$_P$/S$_I$) as function 
of the off axis value is shown in Figure~\ref{smearing}. As shown in 
Figure~\ref{smearing} the bandwidth smearing is negligible up to 
r$\simeq$30 arcmin from the center of each single pointing. Therefore 
the resulting mosaic maps, where the distance between the single pointings 
is $\sim$ 22 arcmin, are not affected by relevant bandwidth smearing and 
point sources should have S$_P$/S$_I>$0.98 everywhere.

\section{The source counts}

\begin{figure}
\centerline{\psfig{figure=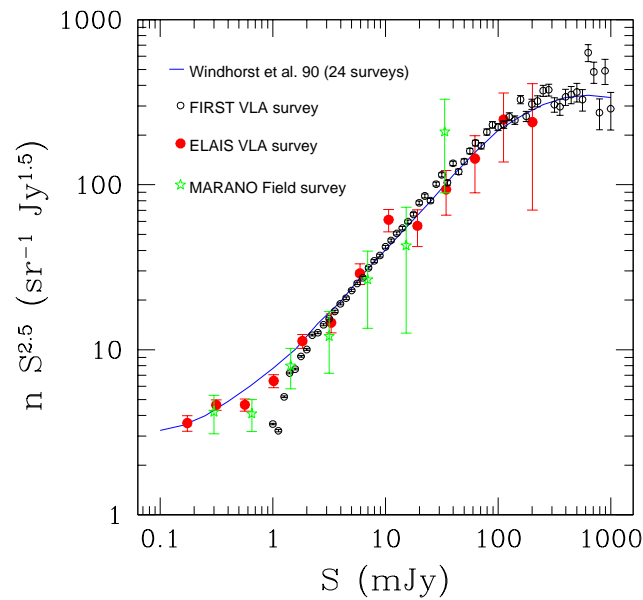,width=9cm}}
\caption[The normalized differential source counts]
{The normalized differential source counts. The abscissa gives the 
flux density (mJy) and the ordinate gives the differential number 
of sources normalized by $S^{2.5}$ (sr$^{-1}$ Jy$^{+1.5}$). The 
solid line represents the global fit to the counts obtained by 
Windhorst, Mathis and Neuschaufer (1990) by fitting the counts 
from 24 different 1.4 GHz surveys. The open circles represent
the counts obtained using the 1.4 GHz survey $FIRST$ (White et al 1997), 
the filled circles are the counts obtained using our VLA observations 
in the ELAIS regions while the open stars represent the 1.4 GHz counts 
obtained by Gruppioni et al. (1997) in the Marano Field. }

\label{counts_ps}

\end{figure}

\begin{figure}
\centerline{\psfig{figure=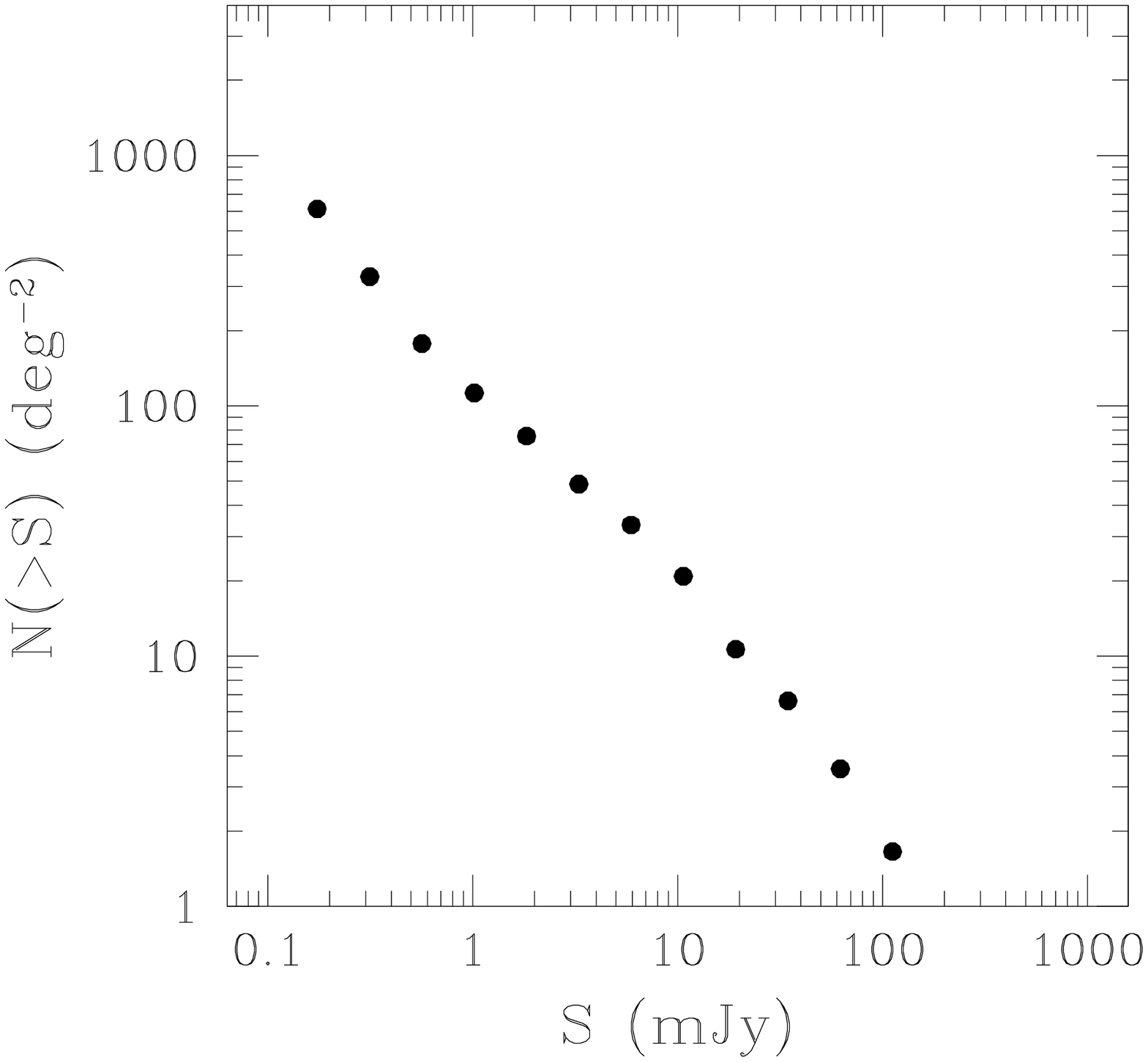,width=9cm}}
\caption{The integral source counts. The abscissa gives the 
flux density (mJy) and the ordinate gives the total number 
of sources per square degree}
\label{counts_int_ps}

\end{figure}

The complete sample of 867 sources with  $S_{peak} > 5\sigma$ was used for 
the construction of the sources counts. Sources with multiple components 
are treated as  single radio sources. For all the sources we have used the 
total flux in computing the source counts. Every source was weighted 
for the reciprocal of its visibility area ($\Omega(S_p)^{-1}$, see 
Figure~\ref{area}) that is the area over which the source could have 
been seen above the adopted limit. The 1.4 GHz source counts of our 
survey are summarized in  Table~\ref{counts_tab}. 
For each flux density bin, the 
average flux density in each interval, the observed number of 
sources in each flux interval, the differential source density
(in sr$^{-1}$ Jy$^{-1}$), the normalized differential counts $nS^{2.5}$
(in sr$^{-1}$ Jy$^{1.5}$) with estimated errors (as $n^{1/2}S^{2.5}$)
and the integral counts (in sr$^{-1}$) are given. The normalized 
1.4 GHz counts 
are plotted in Figure~\ref{counts_ps} where, for comparison, the 
differential source counts obtained with other 1.4 GHz radio surveys 
are also plotted while the integral source counts (deg$^{-2}$) are \
plotted in Figure~\ref{counts_int_ps}. The solid line 
in Figure~\ref{counts_ps} represents the global 
fit to the counts obtained by 
Windhorst, Mathis \& Neuschaufer (1990) by fitting the counts 
from 24 different 1.4 GHz surveys. The open circles represent
the counts obtained using the 1.4 GHz survey $FIRST$ (White et al 1997).  
Finally the open stars represent the 1.4 GHz counts 
obtained by Gruppioni et al. (1997) in a recent radio 
survey in the Marano Field (RA=03$^h$ 15$^m$, DEC=$-$55 13).  

As shown in Figure, there is a very good agreement between our counts 
and those obtained with other surveys. In particular, the points at the 
fainter flux level confirm the well-know flattening observed in the 
normalized differential source counts below few mJy 
(Windhorst, Mathis \& Neuschaufer (1990)). 
The roll-off of the sources in the $FIRST$ survey (open circles
in Figure~\ref{counts_ps}) at flux densities less than $\sim$2 mJy is
due to the low peak flux densities of many faint, extended sources, 
which make them undetectable in the $FIRST$ survey. As noted by 
White et al. (1997) this is a clear indication of the incompleteness
of the $FIRST$ survey at the faint limit. On the other hand, the very 
good agreement between our counts and those obtained with other 
surveys (also at very faint flux level) and the lack of any roll-off
at faint flux level in our counts indicate that our procedure 
for the source extraction (see above) yields very good results and that
our sample is not affected by incompleteness at the faint limit. 

A Maximum Likelihood fit to our 1.4 GHz counts with  two power laws: 
\begin{equation}
\frac{dN}{dS}\propto \left\{ \begin{array}{ll}
S^{-\alpha_1} & \mbox{if S $>$ S$_b$} \\
S^{-\alpha_2} & \mbox{if S $<$ S$_b$}
\end{array}
\right.
\end{equation}
gives the following best fit parameters: $\alpha_1=1.71\pm0.10$, 
$\alpha_2=2.36\pm0.20$, $S_b \simeq$0.5 mJy. 
These values suggest that the re-steepening of the integral counts toward
an Euclidean slope start below 1 mJy, in agreement with Gruppioni et al. 
(1997), Condon \& Mitchell (1984) and Windhorst, van Heerde \& Katgert (1984),
while Windhorst, Mathis \& Neuschaufer (1990), by fitting the counts of 
several 1.4 GHz survey, found that the change in slope starts around 5 mJy. 

\begin{table*}
 \centering
  \begin{minipage}{100mm}
  \caption{The 1.4 GHz Radio Source Counts}
  \label{counts_tab}

\begin{tabular}{cccccc} 
& & & & \\ \hline 
& & & & \\
$S$  & $<S>$ & $N_S$ & $dN/dS$             & $nS^{2.5}$          & $N(>S)$   \\
(mJy) & (mJy) &       & sr$^{-1}$ Jy$^{-1}$ & sr$^{-1}$ Jy$^{1.5}$ & sr$^{-1}$ \\
& & & & \\ \hline
& & & & \\
 ~~0.13 -- ~~0.23 &  ~~0.17  &  87 & 8.962$\times10^9$  & ~~3.60$\pm$~~0.39  &  2.001$\times10^6$ \\
 ~~0.23 -- ~~0.42 &  ~~0.32  & 188 & 2.641$\times10^9$  & ~~4.61$\pm$~~0.34  &  1.076$\times10^6$ \\
 ~~0.42 -- ~~0.76 &  ~~0.56  & 148 & 6.301$\times10^8$  & ~~4.78$\pm$~~0.39  &  5.814$\times10^5$ \\
 ~~0.76 -- ~~1.36 &  ~~1.02  & 127 & 1.990$\times10^8$  & ~~6.56$\pm$~~0.58  &  3.691$\times10^5$ \\
 ~~1.36 -- ~~2.46 &  ~~1.83  & 112 & 8.139$\times10^7$  & ~11.68$\pm$~~1.10  &  2.485$\times10^5$ \\
 ~~2.46 -- ~~4.42 &  ~~3.30  &  64 & 2.543$\times10^7$  & ~15.86$\pm$~~1.98  &  1.597$\times10^5$ \\
 ~~4.42 -- ~~7.96 &  ~~5.93  &  53 & 1.166$\times10^7$  & ~31.61$\pm$~~4.34  &  1.097$\times10^5$ \\
 ~~7.96 -- ~14.33 &  ~10.68  &  43 & 5.251$\times10^6$  & ~61.87$\pm$~~9.43  &  6.842$\times10^4$ \\ 
 ~14.33 -- ~25.79 &  ~19.22  &  17 & 1.153$\times10^6$  & ~59.07$\pm$~14.33  &  3.499$\times10^4$ \\
 ~25.79 -- ~46.41 &  ~34.60  &  13 & 4.900$\times10^5$  & 109.10$\pm$~30.25  &  2.177$\times10^4$ \\
 ~46.42 -- ~83.55 &  ~62.27  &   8 & 1.675$\times10^5$  & 162.10$\pm$~57.31  &  1.166$\times10^4$ \\
 ~83.55 -- 150.39 &  112.09  &   5 & 5.816$\times10^4$  & 244.70$\pm$109.40  &  5.443$\times10^3$ \\
 150.39 -- 270.70 &  201.77  &   2 & 1.293$\times10^4$  & 236.40$\pm$167.10  &  1.555$\times10^3$ \\
& & & & \\ \hline

\end{tabular}
\end{minipage} 
\end{table*}

\section{ Comparison with the $NVSS$ and $FIRST$ radio catalogues}

The regions that we are observing with the VLA are covered also by 
the $NVSS$ and $FIRST$ radio survey. 
The $NVSS$ (NRAO VLA Sky Surveys) covers the sky
north of J2000 $\delta$=$-40^{\circ}$at 1.4 GHz with
$\theta$=45$^{\prime \prime}$ resolution and nearly uniform
sensitivity of 0.45 mJy (1 $\sigma$), while the $FIRST$ (Faint Images
of the Radio Sky at Twenty cm) covers over 10,000 squares degrees at
1.4 GHz with a typical 1 $\sigma$ rms of 0.15 mJy and a resolution of
5.4$^{\prime \prime}$.

It was therefore natural to make a comparison between our results and the 
results obtained by the $NVSS$ and $FIRST$ surveys on the same regions 
of the sky. 

\subsection{SAD on the $NVSS$ and $FIRST$ maps}

As first step we used the $NVSS$ and $FIRST$ radio maps to test 
the reliability of the software that we are using to extract 
the radio sources ({\tt SAD} in the AIPS version October 96). 
From the $NVSS$ and $FIRST$ archive we retrieved the maps covering
the region of the sky in the field N2 covered also by 
the pointing N2 V1, N2 V2 and N2 V3 of  our 
VLA observations (see Table~\ref{t_vla_observations}). 

From the $NVSS$ archive we retrieved the map 
I1640P40 while from the $FIRST$ archive we retrieved the maps 
F16330+40417,  F16330+41132, F16360+40417, 
F16360+41132, F16390+40417 and F16390+41132. 

\begin{figure}
\centerline{\psfig{figure=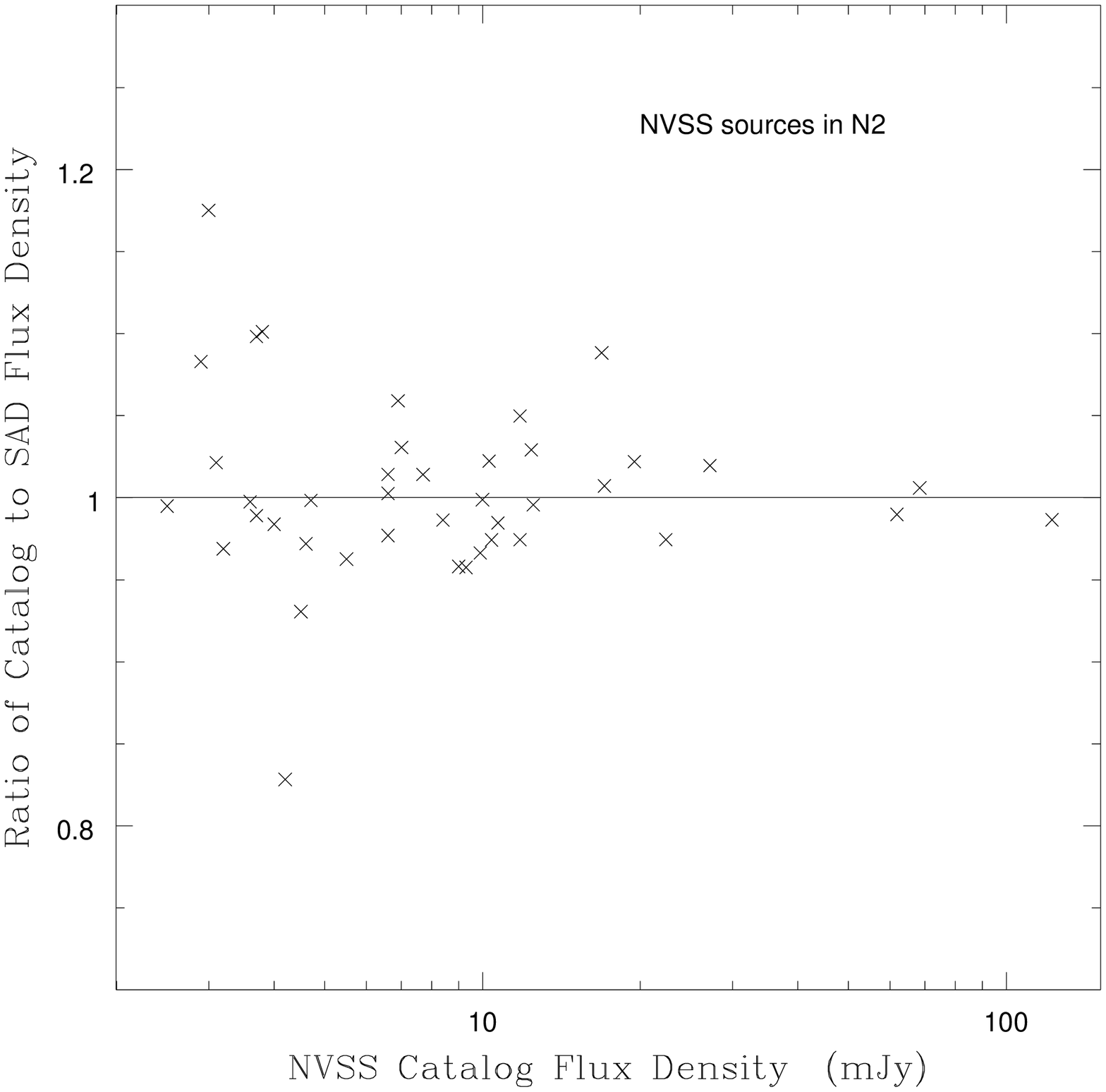,width=8cm} }
\centerline{\psfig{figure=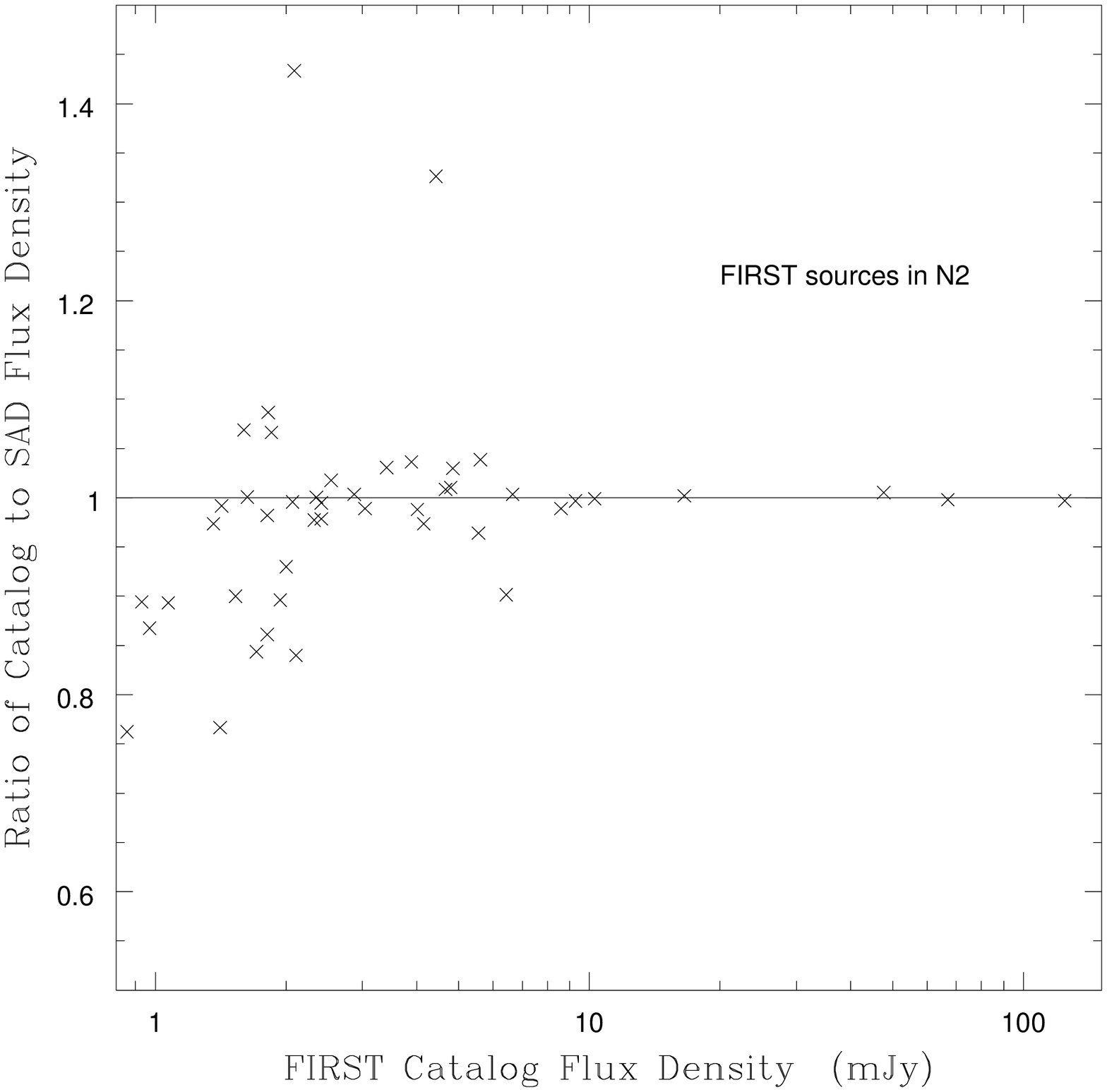,width=8cm} }
\caption{Ratio of flux reported in the public catalogue to  our 
flux obtained using SAD on the radio map 
as function of the flux reported in the catalogue for the $NVSS$  
(top panel) and for the $FIRST$  (lower panel) surveys}
\label{sad_catalog}
\end{figure}

Using the software {\tt SAD} on these maps, we obtained two lists
of sources (one for the $NVSS$ and one for the $FIRST$) in the same 
region covered by our VLA observations. These two lists were 
compared with the list of sources  obtained from the $NVSS$ and 
$FIRST$ catalogs in which are reported the sources above the 
$5\sigma$ limits of each survey. 
The results are reported in Table ~\ref{t_sad} 
and in Figure ~\ref{sad_catalog}. 

\begin{table*}
 \begin{minipage}{100mm}
  \centering
  \caption{Comparison between SAD and $NVSS$ -- $FIRST$ catalogs}
  \label{t_sad}

\begin{tabular}{lccccc} 
& & & & \\ \hline
SURVEY &  Catalogue& {\tt SAD} & Common    & Catalogue          & {\tt SAD} Sources  \\
       &  Sources & Sources   & Sources   & Sources           & not in Catalogue\\
       &          &           &           & not in {\tt SAD}  &     \\
& & & & \\ \hline
$NVSS$  & 46 & 43 & 43 & 3 & 0 \\
$FIRST$ & 74 & 70 & 66 & 8 & 4 \\
& & & & \\ \hline

\end{tabular}
\end{minipage}
\end{table*}

All the three $NVSS$ sources missing in the list obtained by us with {\tt SAD} 
have a peak flux density (calculated by us with {\tt MAXFIT} on the $NVSS$ map) 
lower than $\sim$ 2.0 
mJy ($i.e$ $<$ the 5 $\sigma$ limit of the $NVSS$ maps). Therefore they are 
probably spurious sources. \\
Three (163637.9+410511, 163707.5+405125 and 163815.4+405840) of the eight 
sources present in the $FIRST$ catalogue but not detected by us using {\tt SAD} 
on the $FIRST$ map are not 
detected also in our deeper VLA maps to a flux limit of 0.1 mJy, 
so they are spurious sources. The other five sources 
missing in the $FIRST$ list (as well as the four sources detected by  {\tt SAD} but 
not in the $FIRST$ catalog) have a peak flux density $\lsimeq$1.2 mJy, i.e. very 
close to the $FIRST$ flux limit. These differences are probably due to the 
unreliable Gaussian fit of faint sources as discussed above.
Moreover we have to consider that the sources lists (our  {\tt SAD}
list and the catalogue lists) are obtained using different
source extraction software. In fact, both  $NVSS$ and $FIRST$ sources are extracted 
with special AIPS-based routines wrote exclusively for each survey 
(see Condon et al. 1993 and White et al. 1997 for more details). 

Therefore, in conclusion, we are confident in using
{\tt SAD} as source extraction software, keeping in mind that for 
faint sources the flux obtained with a Gaussian fit may be unreliable.

\subsection{Comparison between our VLA observations and the  
$NVSS$ and $FIRST$ source catalogues} 

The primary difficult in comparing 
the three surveys is the difference in angular 
resolution.  A summary of the 
main properties of the NVSS, FIRST and our VLA survey is reported in 
Table~\ref{t_radio_comparisons}.

The high resolution surveys will partially resolve extended sources 
that appear point--like in the low resolution surveys, so that the flux 
density of the sources will be smaller in $FIRST$ and our survey 
than in the $NVSS$. To minimize this problem, we restricted the comparison 
to compact sources ($i.e$ sources with S$_I$/S$_P<$2) with an off-axis 
value in our VLA map lower than 34 arcmin. The restriction on the 
off-axis value allow us to use only the sources detected in regions 
where the 5$\sigma$ limit is well below the limit of the $FIRST$ survey 
(see Table~\ref{zone_tab} and Table~\ref{t_radio_comparisons}). 
For the $NVSS$ and $FIRST$ surveys we used the public source catalogues as 
available on February 1, 1998.

\begin{table}
 \centering
  \caption{Sensitivity Comparison NVSS, FIRST and our VLA survey}
   \label{t_radio_comparisons}
\begin{tabular}{lccc} 
\\ \hline
                     &  NVSS  &  FIRST  & Our VLA   \\
                     &        &         &  data     \\ \hline
& & & \\
VLA Configuration    &   D    &   B     & C         \\
Beam size             & 45$^{\prime \prime}$ & 5$^{\prime \prime}$&15$^{\prime \prime}$ \\  
confusion noise      & 0.08 mJy  & 0.002 mJy & 0.011 mJy \\
\\
Exposure time        & 30$-$60 s & 3 min & $\sim$1 hour \\
\\
1$\sigma$ rms noise  & 0.45 mJy  & 0.15 mJy & 0.05 mJy \\
5$\sigma$ rms limit  & 2.25 mJy  & 0.75 mJy & 0.25 mJy \\

& & &  \\ \hline
\end{tabular}
Note: The rms confusion from faint extragalactic source in C-configuration
at 20cm is 0.01mJy (Mitchell \& Condon, 1985). 
The value for D-configuration comes from Condon et al. 1998.

\end{table}

\begin{figure}
\centerline{\psfig{figure=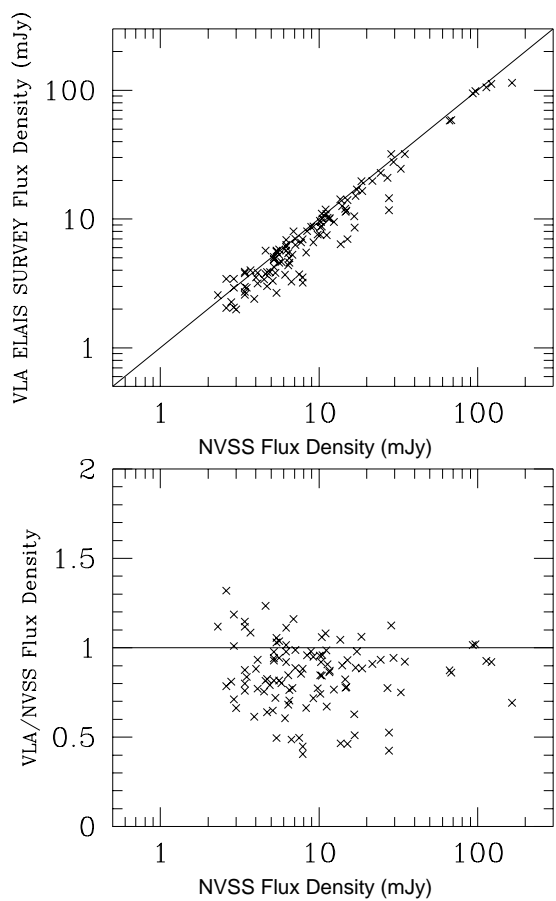,width=12cm}}
\caption{The flux obtained with our VLA survey vs. the  $NVSS$ flux density 
as reported in the public catalogue (upper panel) and the ratio of our VLA
 to $NVSS$ flux density as function of the $NVSS$ flux density (lower panel). }
\label{ff_NVSS}
\end{figure}

\begin{figure}
\centerline{\psfig{figure=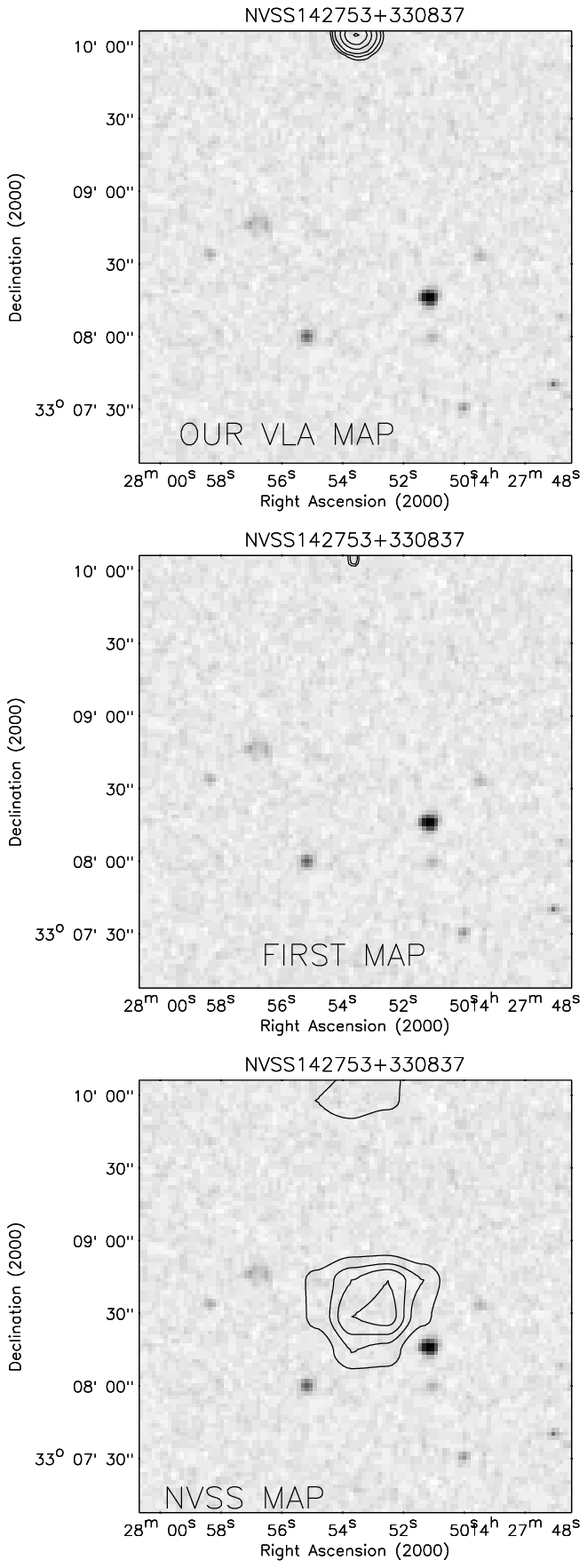,width=15cm}}
\caption{Radio contour plot of the $NVSS$ source NVSS 142753 +330837 present 
in the NVSS catalogue but not in our VLA and $FIRST$ catalogue overimposed to 
the optical Digitized Sky Survey (DSS) image. 
All the maps have a size of 3$^{\prime} \times 3^{\prime}$. 
Radio contour level are 0.25,0.35,0.5,0.75,1.0,1.5,1.8,2.5 mJy for our map, 
0.75,1.0,1.5,1.8,2.5 mJy for the $FIRST$ maps and 
1.0,1.3,1.5,1.7,1.8,2.0,2.2,2.5 mJy for the $NVSS$ map }
\label{nvss_no_vla}
\end{figure}

\subsubsection{Our VLA survey versus the $NVSS$ survey}

\begin{figure*}
%\begin{minipage}{160mm}
%\centerline{\psfig{figure=vla_no_nvss.ps,hight=25cm}}
\centerline{\psfig{figure=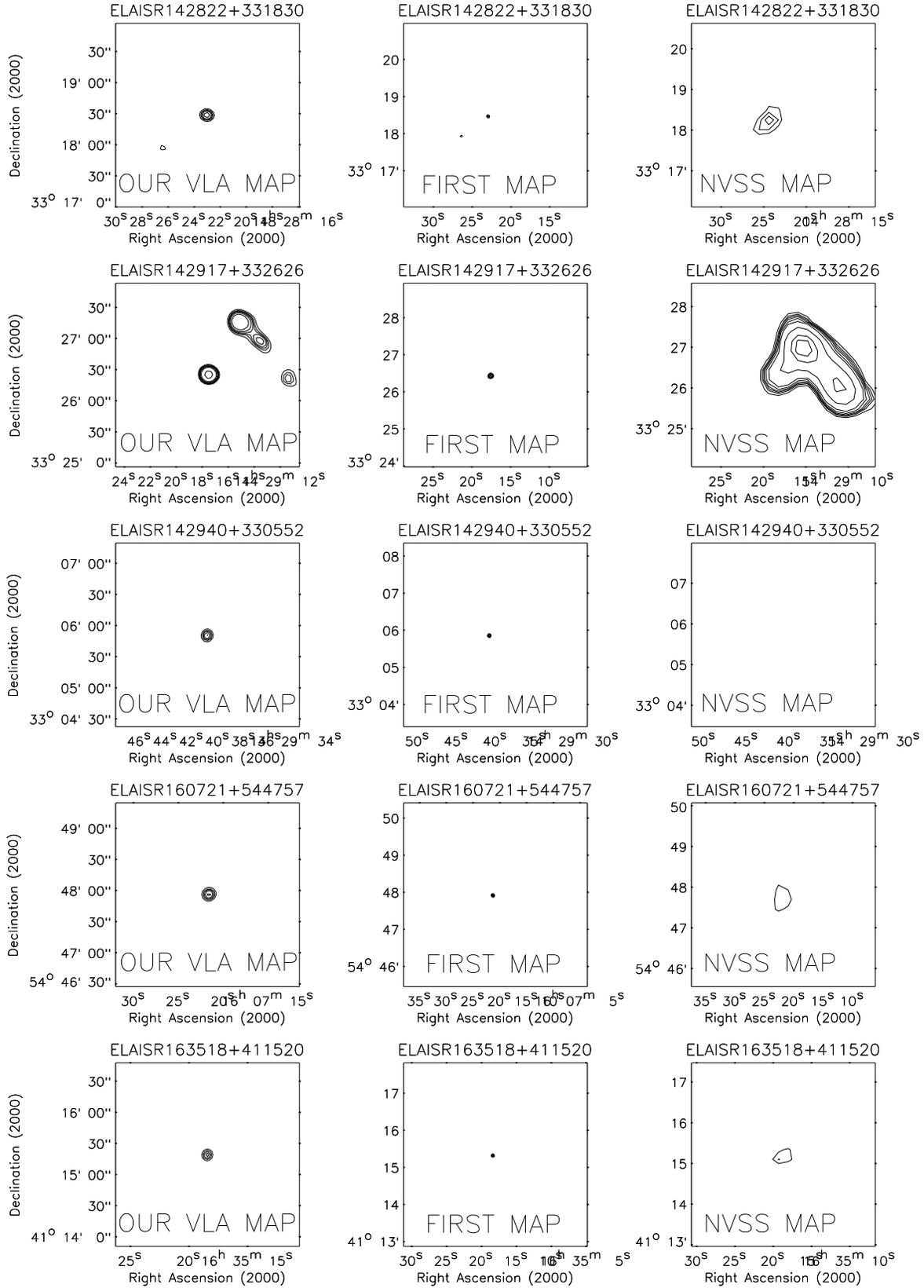,width=22cm}}
\caption{Contour plot of the 5 sources detected in our VLA and $FIRST$ 
survey but not in the $NVSS$ survey. All the maps have a size of 
3$^{\prime} \times 3^{\prime}$ and a the contour 
levels of 1.5,1.8,2.0,2.2,2.5,3,5,8 mJy. }
\label{vla_no_nvss}
%\end{minipage}
\end{figure*}

Using the restriction of an off-axis value lower than 34 arcmin and a 
peak flux density S$_P>$2.3 mJy ($\sim$ the 5$\sigma$ rms limit of the NVSS
survey) we have 109 compact sources in common with the NVSS survey. 
In Figure~\ref{ff_NVSS} we show a comparison between our VLA and 
$NVSS$ total flux densities. A comparison between the flux scales is 
discussed below. Besides these 109 common sources, there is one $NVSS$ source
(NVSS142753+330837) that we did not detected in our survey plus 5 sources  
(with S$_P>$2.3 mJy) detected in our maps but not in the $NVSS$ survey. 
In Figure~\ref{nvss_no_vla} and \ref{vla_no_nvss} we report a contour 
plot of these sources. For comparison a contour plot of the $FIRST$ image in 
the same region of the sky is also reported. 
As shown in Figure~\ref{nvss_no_vla} the $NVSS$ sources missing in our 
survey is not detected also in the $FIRST$ survey. Its $NVSS$ radio flux 
is 2.3 mJy, very close to the 5 $\sigma$ detection limit of the survey. 
It is probably a spurious source but it could be also 
a very low surface brighteness source detected only in the 
$NVSS$, the radio survey with the lowest resolution. However 
a strange phenomenon like a variable radio source can not 
be completely excluded. For the 5 sources 
missing in the $NVSS$ survey, Figure~\ref{vla_no_nvss} clearly shows that 
one (ELAISR142917+332626) is missing because our higher resolution survey 
has resolved an extended $NVSS$ source, one (ELAISR142940+330552) is 
completely undetected on the $NVSS$ map while the other 3  sources are 
detected 
also in the $NVSS$ map but below the 5 $\sigma$ limit ($\sim$ 2.25 mJy). 

\newpage

\newpage

All these five sources are present in the $FIRST$ catalogue (and in the 
$FIRST$ map as clearly showed in Figure~\ref{vla_no_nvss}). Their absence 
in the $NVSS$ catalogue is an indication of the incompleteness of the 
$NVSS$ survey near the flux limit.

\subsubsection{Our VLA survey versus the $FIRST$ survey}

\begin{figure}
\centerline{\psfig{figure=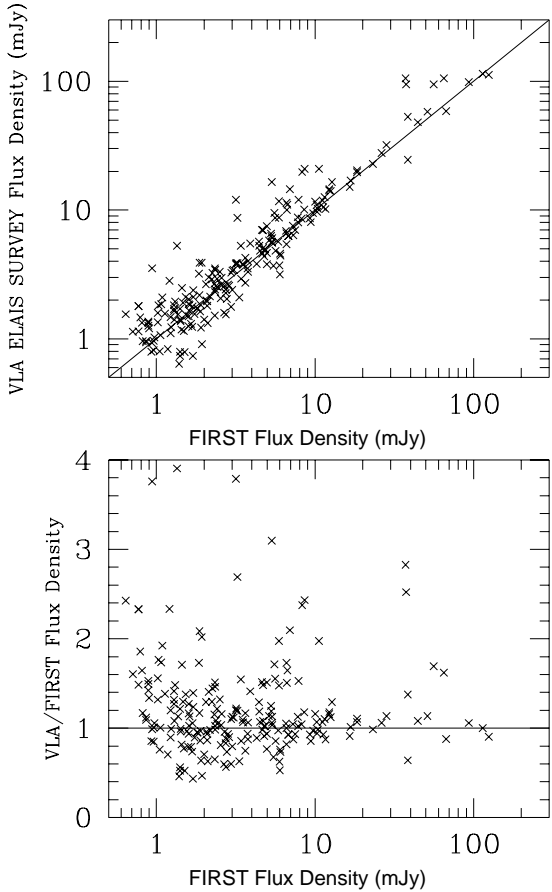,width=12cm}}
\caption{The flux obtained with our VLA survey vs. the  $FIRST$ flux density 
as reported in the public catalogue (upper panel) and the ratio of our VLA
to $FIRST$ flux density as function of the $FIRST$ flux density (lower panel).}
\label{ff_FIRST}
\end{figure}

Using the restriction of an off-axis value lower than 34 arcmin and a 
peak flux density S$_P>$1.0 (the limit of the $FIRST$ catalogue), we have 
215 compact sources in common with the $FIRST$ survey. In 
Figure~\ref{ff_FIRST} we show a comparison between our VLA and 
$FIRST$ total flux densities. 

Besides these 215 common sources, there are 14 $FIRST$ sources that we did 
not detected in our survey and 29 sources present in our survey but not 
in the $FIRST$ catalogue. A flux distribution and a contour maps of these 
last 29 sources are shown in Figure~\ref{vla_no_first_histo} and 
~\ref{vla_no_first} while a contour maps of the 14 $FIRST$ sources that we did 
not detect are shown in Figure ~\ref{first_no_vla}. 
As shown in Figure~\ref{vla_no_first_histo} all  the radio sources 
missing in the $FIRST$ catalogue have a  peak flux density lower than 2 
mJy. However, as shown in Figure~\ref{vla_no_first}, many of them appear 
on the $FIRST$ maps. This result confirms the incompleteness of the $FIRST$ 
survey below $\sim$2 mJy, as already indicated by the differential 
source counts reported in Figure ~\ref{counts_tab}. On the other hand, 
Figure~\ref{first_no_vla} shows that many of the 14 $FIRST$ sources 
missing in our survey  probably are not real. Some of them may be uncleaned 
residual around strong sources (see, for example, FIRST161318+541607, 
FIRST163646+405442, FIRST163815+405840) or simply spurious sources 
in a not well cleaned map (see, for example,  the $FIRST$ maps of 
FIRST161351+543258, FIRST163634+413013 and FIRST163706+405125). 

\begin{figure}
\centerline{\psfig{figure=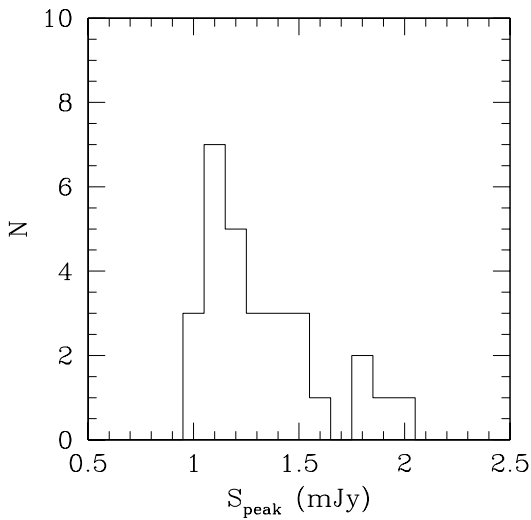,width=8cm}}
\caption{Flux distribution of the 29 sources detected in our survey 
with a peak flux density S$_P>$ 1 mJy but not present in the 
$FIRST$ catalogue}
\label{vla_no_first_histo}
\end{figure}

\subsubsection{Flux comparison between the three surveys}

As shown in Figure~\ref{ff_NVSS} and ~\ref{ff_FIRST} the flux densities 
of our survey are in good agreement with the $FIRST$ and $NVSS$ flux 
densities over two order of magnitude. However, as expected, the high 
resolution surveys tend to estimate lower flux than the lower resolution 
survey. This effect is evident in the lower panels of 
Figure~\ref{ff_NVSS} and ~\ref{ff_FIRST} : our VLA flux densities are, in 
mean, lower than the $NVSS$ flux densities but higher than the $FIRST$ 
flux densities. High resolution surveys, with their smaller synthesized 
beam size, lose flux due to the resolution surface brightness effect. 
However some consideration cam be made from Figure~\ref{ff_NVSS} 
and ~\ref{ff_FIRST}. In spite a factor of 3 in angular resolution between 
our VLA and $NVSS$ survey (15$^{\prime\prime}$ vs. 45$^{\prime\prime}$ FWHM)  
the flux ratio of the two surveys is always lower than 0.5 while the flux 
ratio between our VLA and $FIRST$ surveys reaches values of $\sim$ 4, 
although the two surveys have still a factor of 3 in angular resolution 
differences (5$^{\prime\prime}$ vs. 15$^{\prime\prime}$ FWHM). The missing 
flux between the $FIRST$ (VLA - B configuration) and our survey
(VLA - C configuration) is greater than the missing flux between 
our survey and the $NVSS$ survey (VLA - D configuration). Therefore the C 
configuration of our survey with a synthesized beam size of 
15$^{\prime\prime}$ seems to be the better compromise between high (B 
configuration) and low (D configuration) resolution radio surveys. It is 
less prone to surface brightness effects than the B configuration 
without an excessive loss of flux in comparison with the D configuration. 

\section{Summary}

Using the Very Large Array (VLA) radio telescope, we observed at 1.4 
GHz a total area of 4.222 deg$^2$ in the ISO/ELAIS regions N1 N2 and N3. 
The lower flux density limit reached by our observation is 0.135 mJy 
(at 5 $\sigma$ level) on an area of 0.118  deg$^2$, while the bulk of 
the observed regions are mapped with a flux density limit of 0.250 
mJy (5 $\sigma$). The data were analyzed using the  NRAO {\tt AIPS} 
reduction package. The source extraction has been carried out with 
the {\tt AIPS} task {\tt SAD}. The reliability of  {\tt SAD} has been tested 
using the maps of the radio surveys $FIRST$ and $NVSS$. 

Considering all the available observations, we detected a total of 
867 sources at 5 $\sigma$ level, 44 of which have multiple components. 
These sources were used to calculate the normalized differential source 
counts. They provide a check on catalogue completeness and reliability 
plus information about source evolution. 
A comparison with other surveys shows a very good agreement, 
confirming the presence of the well-know flattening of the counts 
below 1 mJy, the completeness of our catalogue and  the reliability 
of our procedure for the source extraction.

A comparison with the $FIRST$ and $NVSS$ radio surveys has confirmed 
the incompleteness of these two surveys near their flux limits, while a 
flux comparison between the three surveys has shown that our survey 
with the VLA array in C configuration is the best compromise between 
high and low resolution radio surveys. The positional errors of the 
radio sources are $\sim$ 2 arcsec for the fainter sources
($\sim$0.13 mJy) and $\sim$ 0.6  arcsec for the brighter sources ($>$ 10 mJy). 
This small value will enable us to obtain an accurate and fast 
optical/infrared identification of the radio sources. 

\section*{Acknowledgments}
This work was supported by the EC TMR Network programme
(FMRX-CT96-0068). RGM thanks 
the Royal Society for support. 
We thank Rick White for discussions on the optimal pointing
grid and both Bob Becker and Jim Condon 
for discussion on the optimal observing strategy and Bob Becker
for the provision of a FIRST observe file.

\begin{figure}
\centerline{\psfig{figure=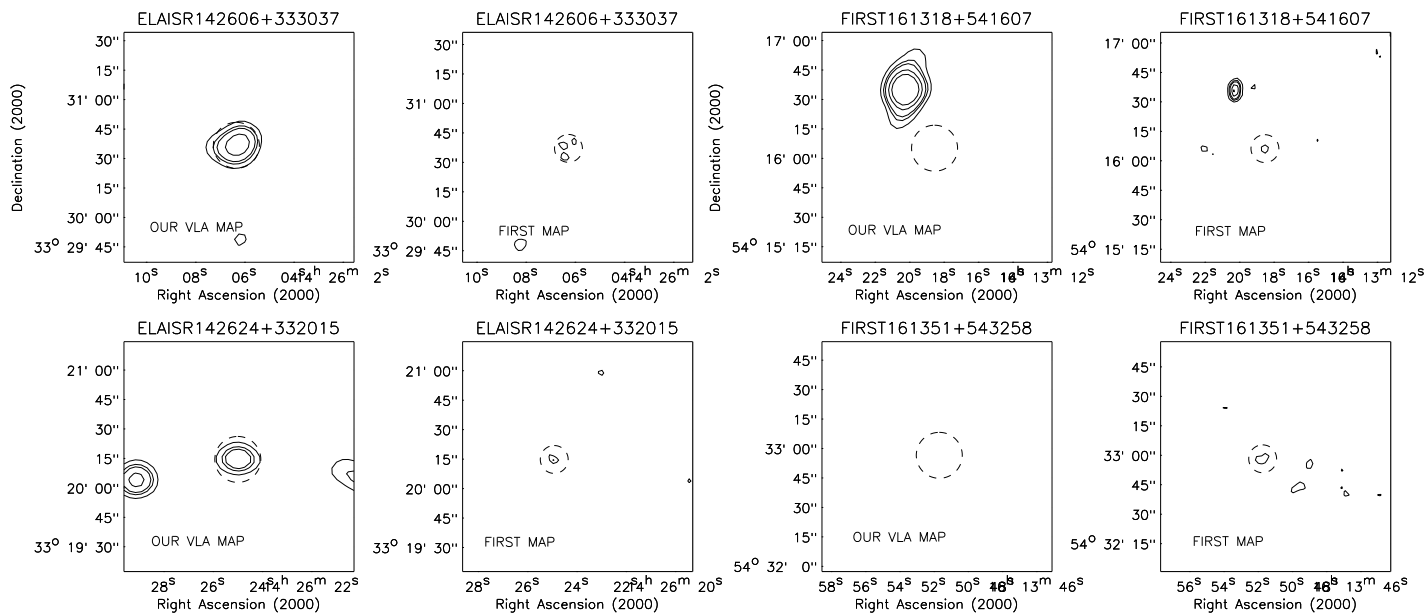,width=9cm}}
%\centerline{\psfig{figure=vla_no_first_1.ps,width=20cm}}
\caption{SMALL EXAMPLE (This figure is available 
in its entirety in http://www.ast.cam.ac.uk/$\sim$ciliegi/elais/papers/)
: Contour plot of the  sources detected in our VLA  
survey with a peak flux density S$_P>$ 1 mJy but not present in the 
$FIRST$ catalogue. The dashed circle shows the position of the source detected 
in our VLA survey. All the maps have a size of 
2$^{\prime} \times 2^{\prime}$ and a  contour 
levels of 0.50,0.8,1,1.5,2,5,8,10 mJy. }
\label{vla_no_first}
\addtocounter{figure}{-1}
\end{figure}

%\begin{figure*}
%\begin{minipage}{160mm}
%\centerline{\psfig{figure=vla_no_first_2.ps}}
%\caption{Continue}
%\addtocounter{figure}{-1}
%\end{minipage}
%\end{figure*}

%\begin{figure*}
%\begin{minipage}{160mm}
%\centerline{\psfig{figure=vla_no_first_3.ps}}
%\caption{Continue}
%\addtocounter{figure}{-1}
%\end{minipage}
%\end{figure*}

%\begin{figure*}
%\begin{minipage}{160mm}
%\centerline{\psfig{figure=vla_no_first_4.ps}}
%\caption{Continue}
%\end{minipage}
%\end{figure*}

\begin{figure}
\centerline{\psfig{figure=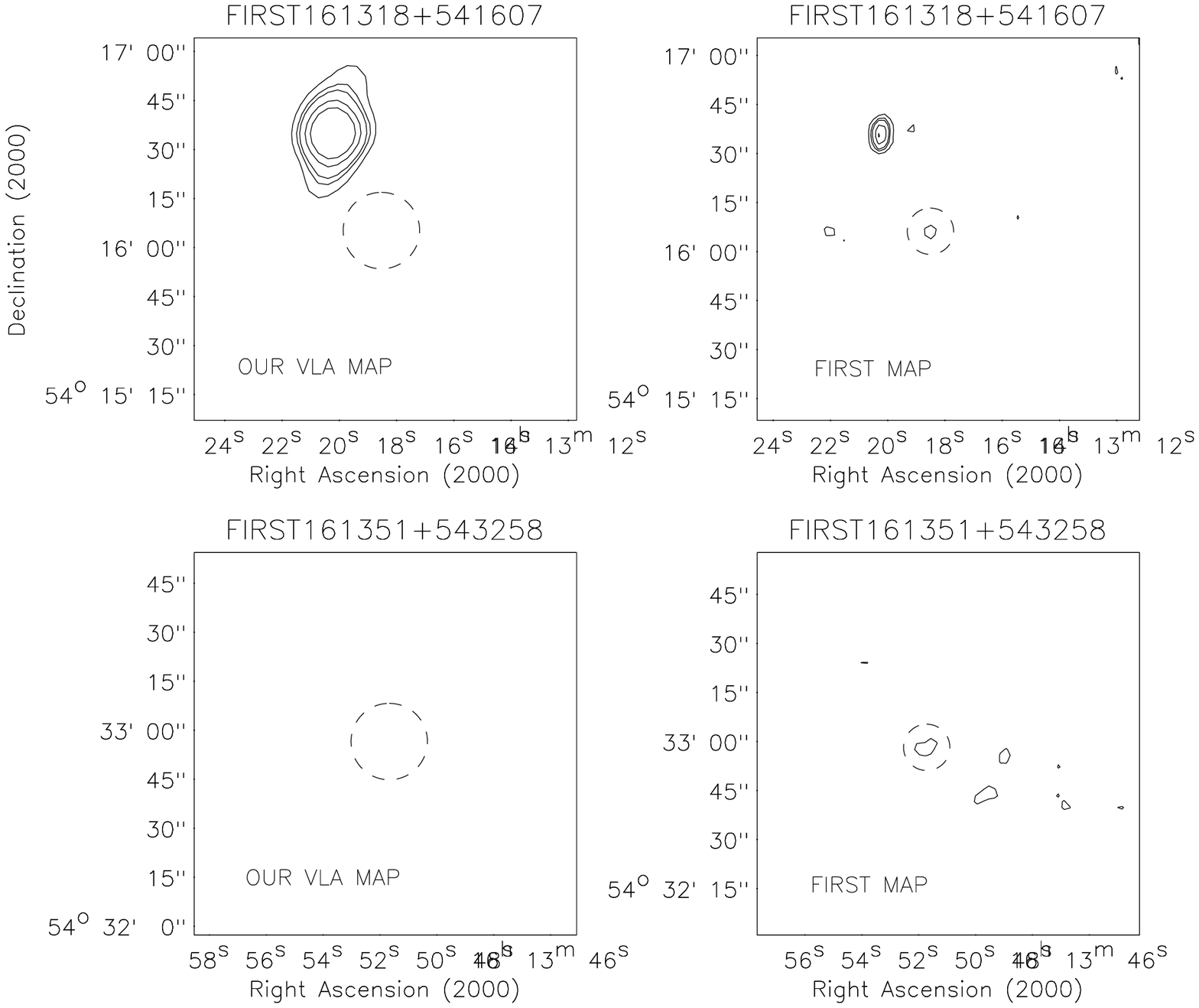,width=9cm}}
\caption{SMALL EXAMPLE (This figure is available 
in its entirety in http://www.ast.cam.ac.uk/$\sim$ciliegi/elais/papers/):
 Contour plot of the $FIRST$ source  not in our VLA 
catalogue. The dashed circle shows the position of the $FIRST$ source.
All the maps have a size of 
2$^{\prime} \times 2^{\prime}$ and a  contour 
levels of 0.50,0.8,1,1.5,2,5,8,10 mJy. }
\addtocounter{figure}{-1}
\label{first_no_vla}
\end{figure}

%\begin{figure*}
%\begin{minipage}{160mm}
%\centerline{\psfig{figure=first_no_vla_2.ps}}
%\centerline{\psfig{figure=vla_no_nvss.ps,width=20cm}}
%\caption{Continue}
%\end{minipage}
%\end{figure*}

\label{lastpage}

\end{document}